\documentclass[preprint,10pt,floatfix,nofootinbib]{revtex4}
\usepackage{graphicx}
\usepackage{amssymb}
\begin{document}

\message{TBD: NEXT TIME: NEED COMPLEX Z AND GROUP DELAY FOR HORNS/YAGIS TO PREDICT EXPECTED SIGNAL SHAPE.}


\message{TBD: COMPARE WITH SOAR DATA AND ALSO VOSTOK DATA}

\message{``WE ATTRIBUTE DIFFERENCES BW THESE WFS AND TD WFS TO SIGNAL GENERATOR DIFFERENCES''}

\message{TBD: 0809: A) YAGIS BURIED IN SNOW B) FROM SPASE SHACK OR SOMEWHERE ELSE AWAY FROM MAPO REFLECTIONS}

\message{
> okay, 90 degrees I understand, assuming that the ordinary and 
> extraordinary axes are orthogonal to each other. How good is this 
> assumption?
>   
It's a reasonable assumption except for great depths close to the bed 
(say 10
complex so that ice fabric there is also complex.  Shuji Fujita and I 
published a paper in JGlac 2006 which shows a reasonable match between 
radar-detected features and ice-core features derived with the 90-degree 
assumption.

> were the 90 degree and the 180 degree modulations clearly distinguishable 
> (i.e. not aligned)?
>   

Yes at many cases.  See my 2004 JGR paper.  I identified significant 
number of clear 90-degree and 180-degree features, although there are 
many non-typical cases however.

> when you say 'depth', you mean ice-thickness to bedrock, is that right? 
> I.e., you assume that the bottom reflection dominates. And this was varied 
> just by changing your locale, is that right?
>   
Depth refers distance from the radar system on the ground to a reflector 
(either bed or internal horizons).

> what was the magnitude of the amplitude difference that you measured for 
> the 'anisotropic reflector' effect? Also, how much power did you observe 
> in the cross-polarization configuration (i.e., power scattered out of the 
> co-pol configuration and into the x-pol configuration).
>   
Azimuthal variations of the echo caused by anisotropic reflector has an 
magnitude larger than 10 dB.
I don't have echo intensity of cross-polarization handy but you can have 
some ideas from papers by Fujita et al. in JGlac 2006.

>   
>> The former can be 
>> caused by anisotropic (biaxial) reflectors too, but it is quite unlikely 
>> in the central of ice sheets (including South Pole).  Multiple 
>> reflection between layers is negligible since reflectivity is about -70 
>> dB so double reflection (reflection up -> reflection down -> reflection 
>> up -> radar system) will be 140 (= 70 x 2) dB weaker than the primary 
>> reflection.  I am afraid that I don't answer your question.  If so, 
>> please feel free to write to me.  I stay at home today because all 
>> others are in the bed due to cold, but I can read emails.
>>
>>     
>
>>> hi kenny-
>>>
>>> I had a couple of questions for you - a) can you remind me what kind of 
>>> antennas you used for your birefringence measurements around dome fuji? b) 
>>> I did some more measurements at pole this year, in which I looked at the 
>>> signal returns from internal scattering layers. I do see a 2pi modulation 
>>> in amplitude, but no time stagger between the reflection returns. If so, 
>>> its not clear to me that what you saw before was not 
>>> polarization-dependent reflection and not birefringence. can you remind me 
>>> how you excluded this possiblity?
>>>
>>> thanks. i hope you're having a good holiday break!
>>> dave
>>>
>>> p.s. no news on the proposal, I'm assuming...}

\message{http://docserver.ingentaconnect.com.www2.lib.ku.edu:2048/deliver/connect/igsoc/00221430/v52n178/s9.pdf?expires=1198871497&id=41445237&titleid=6497&accname=University+of+Kansas&checksum=C772476BFD8738DB1328CC1AB8B20251 - ``For pulse widths between 60 and 1000 ns, which are typically used in ice soundings, the depth range is 5-85 m''}

\title{Polarization dependence of radiowave propagation through
Antarctic ice}
\author{D. Besson} 
\affiliation{Dept. of Physics \& Astronomy, U. of Kansas, 
Lawrence, KS  66045}

\begin{abstract}
Using a bistatic radar system on the ice surface,
we have studied radiofrequency reflections off internal layers in
Antarctic ice at the South Pole. In our measurement,
the total propagation time of
$\sim$ns-duration, vertically broadcast radio signals, as a function
of polarization axis in the horizontal plane, provides a direct
probe of the geometry-dependence of the ice permittivity to depths of 1--2 km.
Previous studies in East Antarctica have interpreted
the measured azimuthal dependence of reflected signals 
as evidence for birefringent-induced interference effects, which
are proposed to result from preferred alignment of the crystal
orientation fabric (COF) axis. To the extent that COF alignment results
from the bulk flow of ice across the 
Antarctic continent, we would expect
a measurable birefringent asymmetry at South Pole, as well.
Although we also observe clear
dependence of reflected amplitude on polarization angle in our
measurements, we do not
observe direct evidence for birefringent-induced time-delay
effects at the level of 0.1 parts per mille. 
\end{abstract}

\maketitle

\section{Introduction}
Efforts are underway to use the Antarctic icecap as a neutrino
target\citep{ICECUBE,RICE,AURA,ANITA}. Neutrino-ice
collisions result in the production of 
charged particles which
emanate from the collision point at $v\to c$.
In a medium with index-of-refraction $n>1$, detection of the resulting
Cherenkov radiation
in either the near-UV or radio wavelength regime by a suite of
sensors can be used to reconstruct the kinematics of the initial
neutrino, provided the 
absorption and refraction of the original electromagnetic
signal due to the intervening ice can be reliably 
estimated. Complete characterization of the ice permittivity,
as a function of depth and also polarization is
therefore important in obtaining a reliable estimate of the
neutrino detection efficiency.

For a radio receiver array,
sensitivity is optimized by probing a large
mass (i.e., large neutrino target) of cold (long attenuation length),
uniform (minimal sensitivity to systematic errors
resulting from local anisotropies)
clean (no scattering) ice. Although surface elevations vary across
the continent, ice thicknesses are typically $\sim$3 km for 
most of the potential radio array sites, as indicated in
Table \ref{thickness-data}.
Thanks to its superb infrastructure, South Polar ice is perhaps
the most extensively characterized at a single site\citep{RF-eps-re,RF-eps-im}. Our
current study represents an extension of previous 
ice dielectric measurements at South Pole.
\begin{table}
\begin{center}
\begin{tabular}{c|c|c|c|c}
  Ice Dome & Coordinates  & Height (m)   & Bed Elevation (m) &  Ice Thickness (m) \\ \hline
  Dome A   & 81 S, 77 E & 4093   & 1597 & 2486 \\
  Dome C   & 75 S, 125 E & 3233   & 249 & 3270 \\
  Dome Fuji   & 77 S, 37 E & 3786   & 963 & 2823 \\
  Vostok   & 77 S, 104 E & 3529   & 352 & 3177 \\
  South Pole   & 90 S & 2771   & --57 & 2828 \\ \hline
\end{tabular}
\caption{Data (taken from the BEDMAP collaboration)}
\label{thickness-data}
\end{center}
\end{table}

The long radiofrequency attenuation length of ice has also facilitated
extensive aerial surveys of the icecap, which have yielded not
only measurements of the ice depth, but also detection of
internal reflecting layers in the ice itself over 
horizontal distance
scales of thousands of kms in both
Greenland\citep{CRESIS} and Antarctica\citep{SOARmain,SOARSP}. Combining radio
reflection data from aerial surveys with site-specific ice
core data, as well as data taken on traverses,
attempts have been made to construct generalized models of the ice
radiofrequency response, as a function of density,
chemistry and crystal-orientation-fabric (COF) geometry.
Perhaps most interesting are polarization-specific asymmetries
in ice dielectric response (including `birefringence').
There have been several measurements made of Antarctic ice which have
been interpreted as 
evidence for 
birefringence\citep{Hargreaves,Hargreaves-1978,Matsuoka-biref,Matsuoka2004,Doake2002,Doake2003}. In a comprehensive
study based on data taken along a traverse in the vicinity of Dome Fuji,
co-polarized 
(transmitter and receiver antenna
polarizations parallel, projected onto the
horizontal plane) 
and cross-polarized 
(transmitter and receiver antenna
polarizations perpendicular, projected onto the
horizontal plane) 
signals were broadcast from a network
analyzer operating at a given
frequency using three-element
Yagis. Modulation of the received amplitude with period $2\pi$ and
$4\pi$ radians were observed. Under the assumption that ordinary and
extra-ordinary birefringent axes are orthogonal,
the latter was interpreted as due to
birefringent-induced interference effects in the time-domain; the
former was interpreted as anisotropic reflections.
In that study, the physical consequences of radar scattering
due to density, COF, and acidity effects were considered;
the possible effects of Faraday rotation were ignored. 
In order to simplify the interpretation of data, the authors
made the important assumption that one scattering
cause was primarily
responsible for all radar polarizations, and further presume that
\message{Table 1 from matsuoka et al 2006}
although all types of scattering can result
in large cross-polarized signals, only COF produces anisotropic scattering
which also results in a time-delay between signals
received along appropriately aligned orthogonal axes.
\message{Not in Table 1, but quote - This is because the effect of PCOF
appears in one orientation but the effects of the other mechanisms
can appear in the orthogonal orientation.}
COF and acidity-based scattering 
give returns which are typically reduced a factor of 1000
in amplitude; density scattering yields returns down a typical factor of
100 in amplitude. 

\begin{figure}
\centerline{\includegraphics[width=4cm]{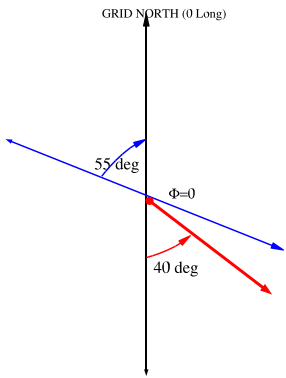}}
\caption{Geometry of measurements presented herein. $\phi$=0 corresponds
to reference for primary reflection data. 
Surface iceflow direction at South Pole is shown in red, corresponding
to $\phi$=--15 degrees ($\equiv$+165 degrees). Magnitude of iceflow
is measured to be 10 m/yr, and uniform to a depth of $\sim$2 km. Measurements
taken at South Pole in January, 2004, referred to later
in this document, corresponded to an antenna alignment
axis of $\sim$--11 degrees (+169 degrees).}
\label{fig:iceflow}
\end{figure}

More recently,
there have been attempts to establish a link between birefringence
and COF alignment, presumably due to the stress of the local ice flow.
In such a model, knowing the ice flow history allows
one to extract the crystal
orientation and therefore predict the birefringence axes.
We previously used the reflection off the bedrock observed at a site
near Taylor Dome to quantify the birefringent asymmetry, projected
onto the vertical ${\hat z}$-axis 
(perpendicular to the surface) of order 0.12\%, although
a correlation with the local ice flow was not fully established\citep{TD-paper}.
Consistent with the predictions of \citep{Matsuoka2004}, some
cross-polarized signal was observed, however, this signal was relatively
small and entirely consistent with the cross-polarization 
cross-talk specifications of
the antennas used. Using a technique similar to that
employed for our Taylor Dome study,
herein we again search for birefringent asymmetries for signals propagating along the
${\hat z}$-axis through the ice.
Our objective
was to use internal scattering layers as multiple reflecting planes to
 calculate birefringent effects as a function of depth,
and ultimately correlate the depth-dependent birefringence with
COF-alignment and ice flow as a function of depth.
Globally, South Polar ice is observed to flow at a 
the surface with a velocity of approximately
10 m/yr in a direction which corresponds to about 160 degrees in the
coordinate system used for these measurements\citep{ice-shear}
(corresponding to the 40 degree West Longitude line).
The geometry of our measurements are shown in 
Figure \ref{fig:iceflow}. 

\section{Set-up, Calibration, and in-air broadcasting}
\begin{figure}
\centerline{\includegraphics[width=7cm]{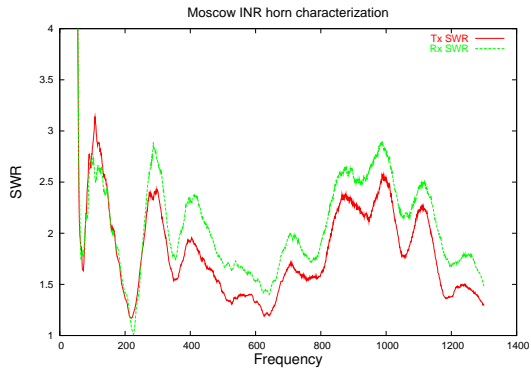}}
\caption{Voltage Standing Wave Ratio (VSWR) for horn antennas used in measurements described herein.}
\label{fig:horn_swr}
\end{figure}

Two 1/2-inch thick jumper coaxial cables, each approximately 25 m long, were fed 
from the RICE experiment inside
the Martin A. Pomerantz Observatory (MAPO)
building at the South Pole through a
conduit at the bottom of the building and out onto the snow. 
Each cable
was then connected to a TEM horn antenna, manufactured by the Institute of
Nuclear Research (INR), Moscow. 
These antennas were also used in our
previous measurement of the ice attenuation length at the South Pole\citep{RF-eps-im}.
As with our previous measurement,
for in-ice transmission, each horn antenna 
is placed face-down on 
the surface
looking into the snow. Signals are lowpass filtered, with a variety of
filters, then high-pass filtered to remove components above 1 GHz,
notch filtered to suppress the large South Pole background 
noise at 450 MHz which serves as the station Land Mobile Radio
carrier, and finally amplified by +52 dB prior to data acquisition.
The horn antennas have reasonably good
transmission characteristics, from 60 MHz up to 1300 MHz, as
indicated by the Voltage Standing Wave Ratio (VSWR) plots in
Fig. \ref{fig:horn_swr}. 
We note that, within 1--2 wavelengths, downward-facing horns see
some mixture of air+snow, which down-shifts the antenna response in frequency
by some 10-20\%, and also narrows the beam pattern. 
The measured frequency response shown in
Figure \ref{fig:horn_swr} is that of the horn antennas in their 
experimental configuration and therefore should correctly represent the
horn characteristics relevant to this measurement. The forward gain
of the horns is expected to be $\sim$10 dBi in air, or 
$\sim$12--15 dBi in-ice.
In an attempt to minimize in-air ringing between the transmitter horn
(Tx) and the
receiver horn
(Rx), the two antennas were placed on opposite sides of MAPO.
For in-air broadcasting measurements made as
a cross-check of polarization isolation, the horns were propped up
onto the snow surface with their open ends facing each other on the same
side of MAPO.
In this case, the +52 dB amplifier was removed and signals were 
read directly at the digital oscilloscope.

\begin{figure}
\centerline{\includegraphics[width=7cm]{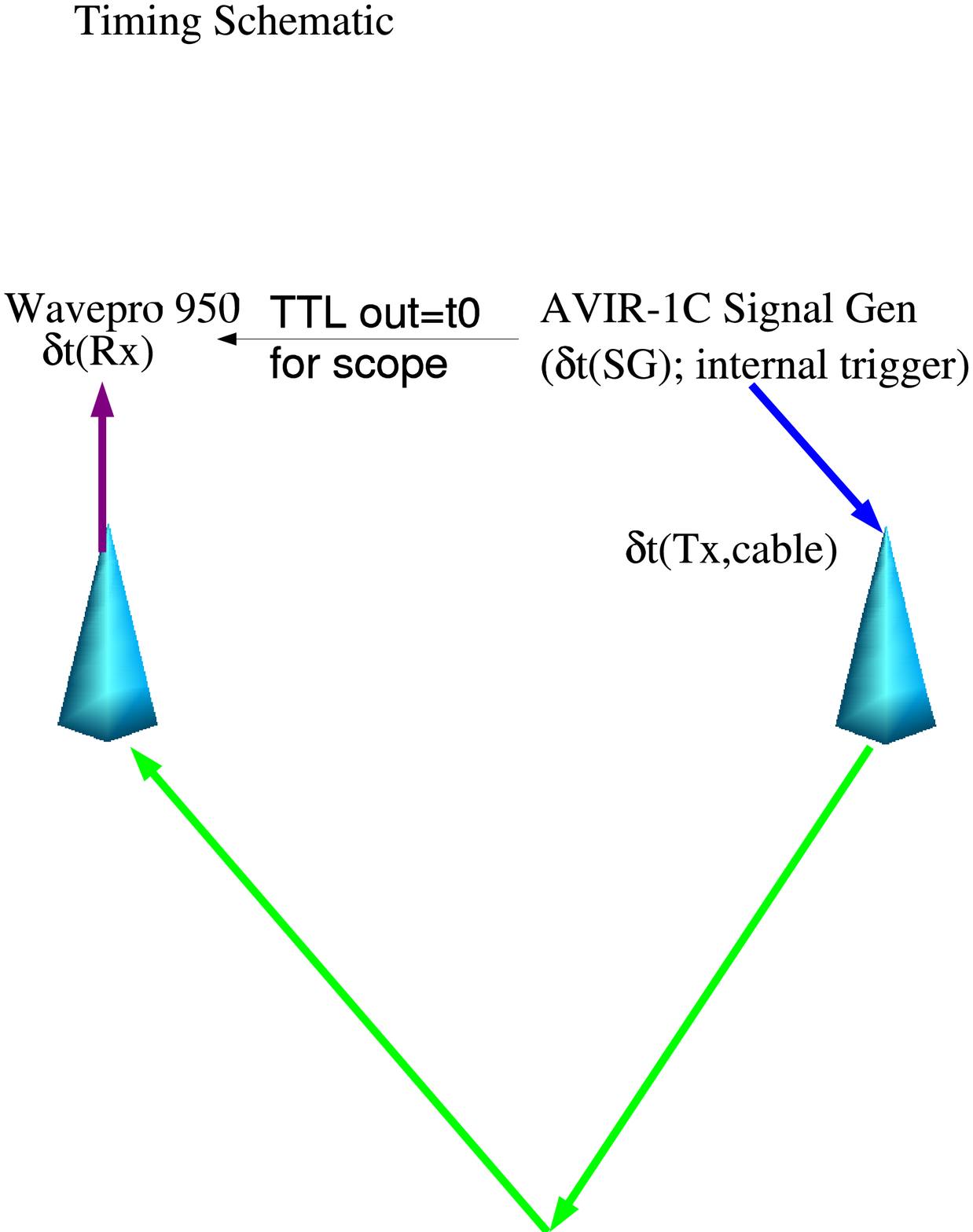}}
\caption{Trigger timing schematic for primary measurements described herein.}
\label{fig:timing_schematic_trig3}
\end{figure}

\begin{figure}
\centerline{\includegraphics[width=10cm]{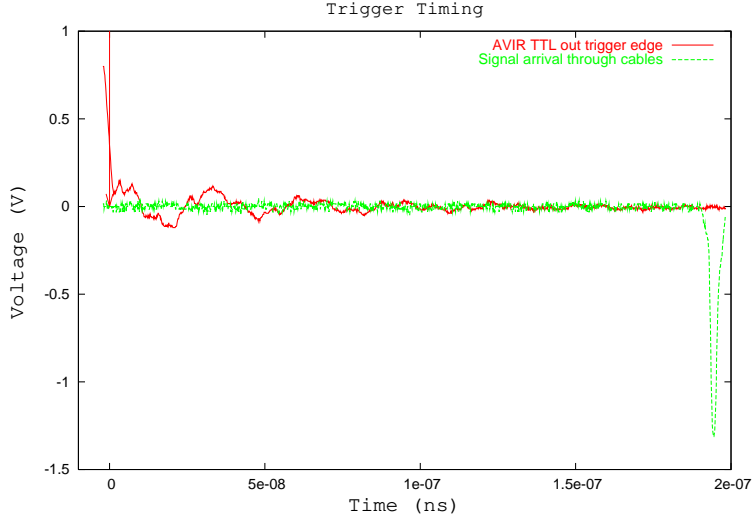}}
\caption{Oscilloscope trigger (falling edge on left) 
with respect to signal through cables  (as in previous plot, with time scale and voltage units more explicit).}
\label{fig:Trigger-timing-check}
\end{figure}

Signals were taken from an AVIR-1C pulse generator into the transmitter
horn; data acquisition of receiver horn 
waveforms was performed using the LeCroy 950 Waverunner
digital oscilloscope. This scope features good bandwidth (1 GHz)
and a high maximum digitization speed (16 GSa/sec). For the
measurements described herein, the scope sampling rate was generally
set to 2 GSa/sec. In order to enhance
signal-to-noise, many waveforms (40,000 typically) were averaged.
As waveform averaging requires a) stable scope capture trigger relative
to b) the signal output, both a) and b) were taken directly from the
AVIR-1C, as depicted in Figure \ref{fig:timing_schematic_trig3}.
In order to correlate our observed internally-reflected signals with
previous measurements, we must subtract off the time-delay due to
cable propagation. We measure a cable propagation time delay of
($\delta t_{cable}$) $\approx$+190 ns (Figure \ref{fig:Trigger-timing-check})
relative to the oscilloscope trigger time.

\begin{figure}
\begin{minipage}{18pc}
\includegraphics[width=7cm]{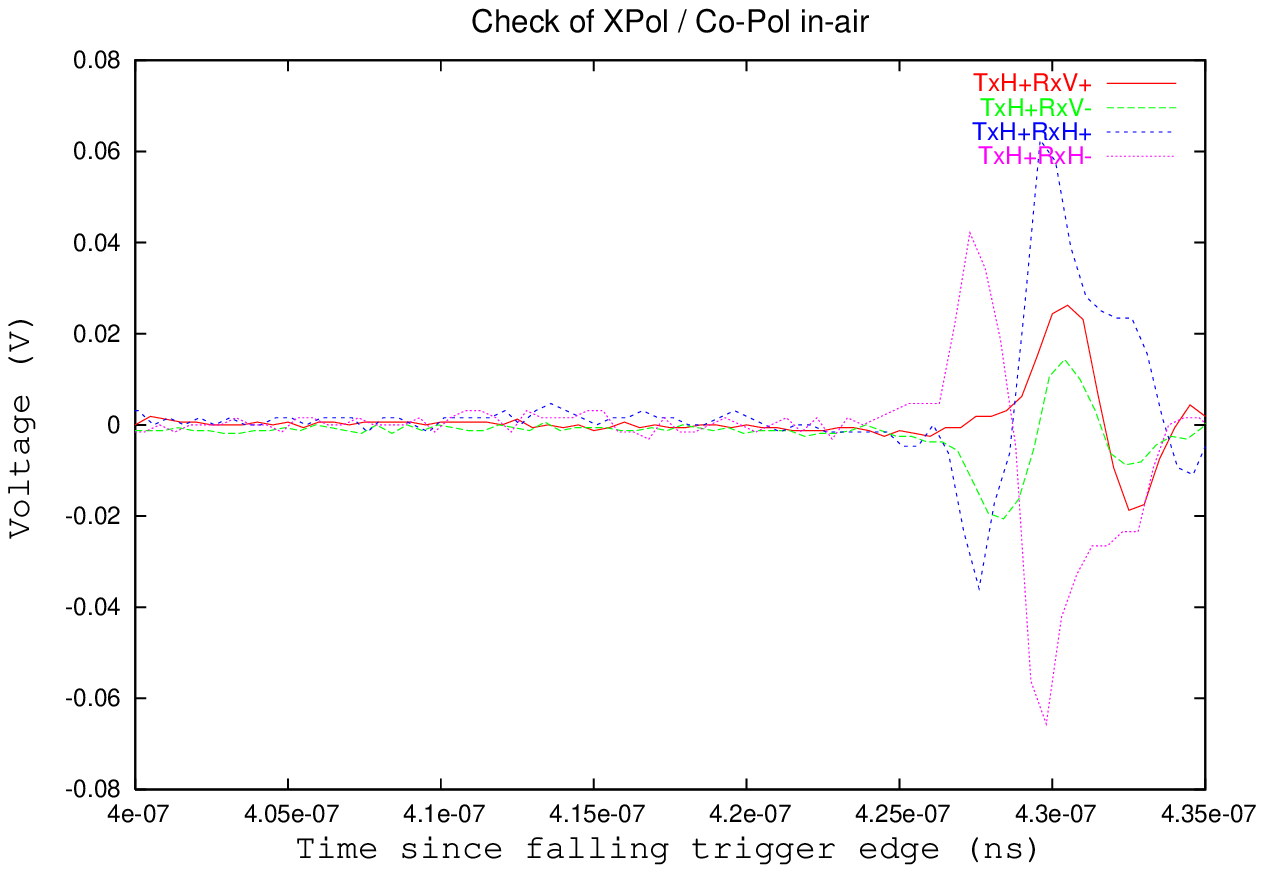}
\caption{In-air broadcasts, 93 foot separation between
transmitter and receiver; transmitter oriented horizontally in
``H+'' configuration.}
\label{fig:inair-d93ft-check-TxH+}
\end{minipage}
\hspace{1pc}
\begin{minipage}{18pc}
\includegraphics[width=7cm]{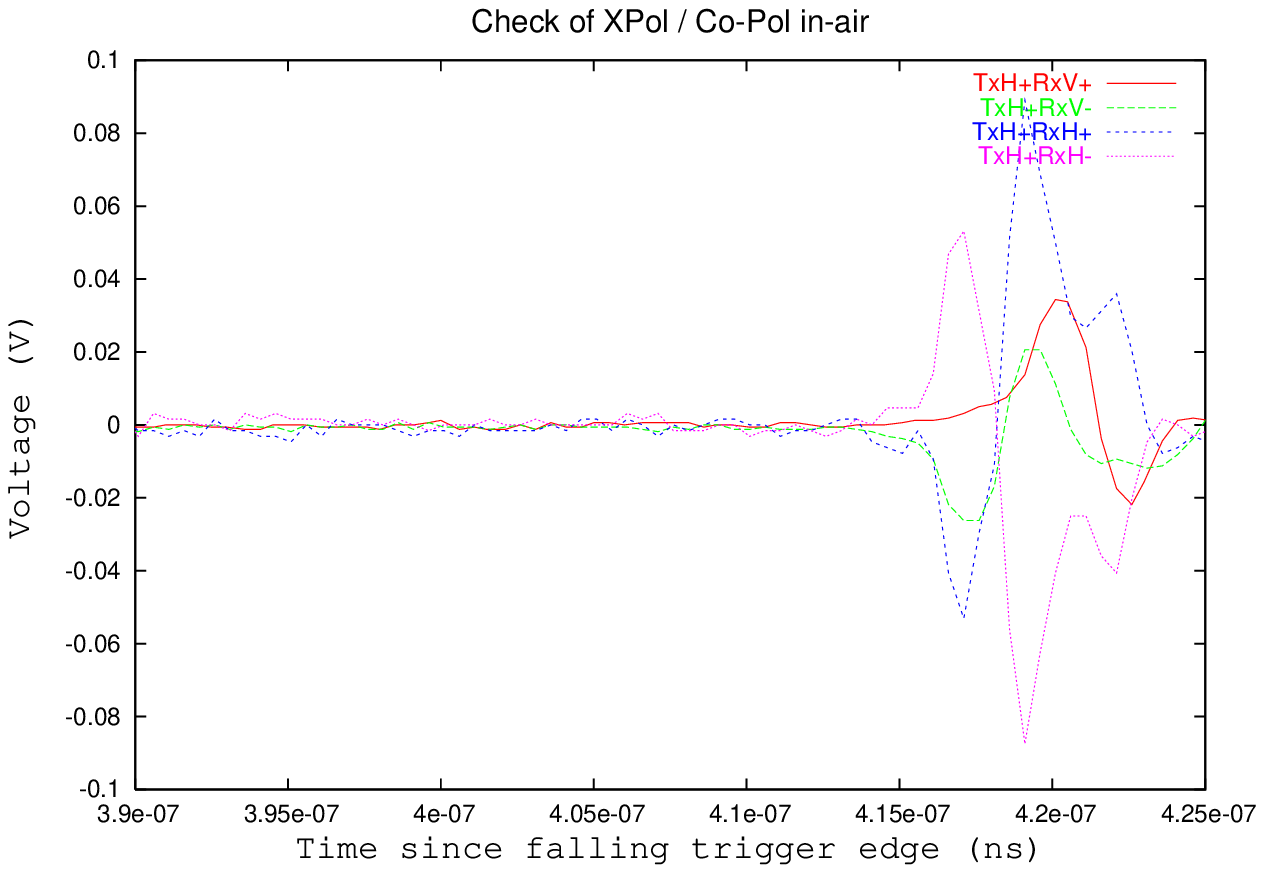}
\caption{In-air broadcasts, 83 foot separation between
transmitter and receiver; transmitter oriented horizontally in
``H+'' configuration.}
\label{fig:inair-d83ft-check-TxH+}
\end{minipage}
\end{figure}

We have attempted to estimate the isolation between ``Vpol'' (along
the long axis of the horns) and ``Hpol'' (the perpendicular axis) by
broadcasting signals in-air, with both antennas on the same
side of MAPO and facing each other on the snow. 
This measurement is complicated by the
possibility of reflections off both the snow surface itself, as well
as the aluminum
side of the MAPO building. Although previous similar work at Taylor Dome
indicates that the former reflection does not seem to produce
substantially noticeable effects, the reflection off the metal
sides of MAPO is likely to be non-negligible.
Because of interference between the three possible routes
(direct plus the two reflections mentioned above), we restrict our
attention to the first few ns of received signal when broadcasting
directly through air. In order to qualitatively assess interference
effects at a particular separation distance,
measurements were made at separation distances
of 93 ft. and 83 ft., respectively. Antenna geometry is referenced relative
to the side of the TEM horn where the cable is
attached to the feed point; this side
is defined as ``+''. We define ``+H'' as the antenna orientation when
the antenna is lying horizontally and the feed cable points directly
away from MAPO; ``+V'' denotes the antenna orientation when the antenna
is oriented vertically with the feed cable pointing away from the
snow surface. Based on Figures
\ref{fig:inair-d93ft-check-TxH+} and
\ref{fig:inair-d83ft-check-TxH+}, 
the estimated V:H power isolation, 
when the transmitter is broadcasting
lying 'horizontally' on the snow is approximately 5 dB. The signal
duration is estimated to be of order 5--10 ns.

\begin{figure}
\centerline{\includegraphics[width=7cm]{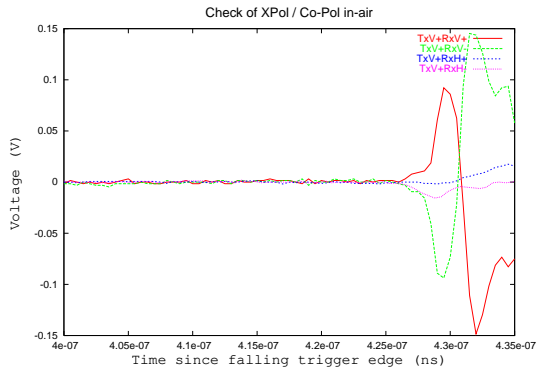}}
\caption{In-air broadcasts, 93 foot separation between
transmitter and receiver; transmitter oriented horizontally in
``V+'' configuration.}
\label{fig:inair-d93ft-check-TxV+}
\end{figure}

When the transmitter
is broadcasting in the ``+V'' configuration, the cross-talk is considerably
smaller (Figure \ref{fig:inair-d93ft-check-TxV+}). Isolation of 
5 dB should be adequate in our search for birefringent effects. 
Based on these Figures, we also estimate the total absolute gain uncertainty
to be approximately a factor of two.
Note
that an antenna with an azimuthally-symmetric monopole beam pattern has
no sensitivity to birefringence. In that case, any arbitrary signal has
the same components along the ordinary and extraordinary axes.

\section{In-Ice broadcasting}
\begin{figure}
\centerline{\includegraphics[width=7cm]{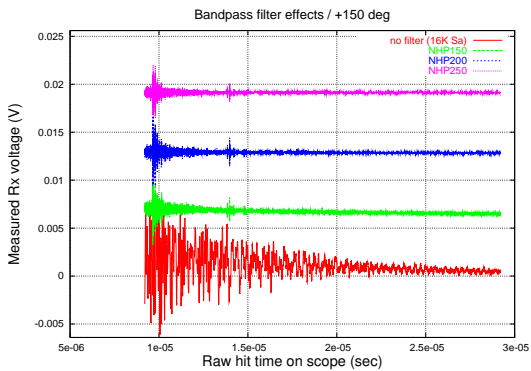}}
\caption{Comparison of received signals, for different filters applied to receiver output. We use, as a default, the NHP250 filter.}
\label{fig:Filter_comparison}
\end{figure}

\begin{figure}
\begin{minipage}{18pc}
\includegraphics[width=7cm]{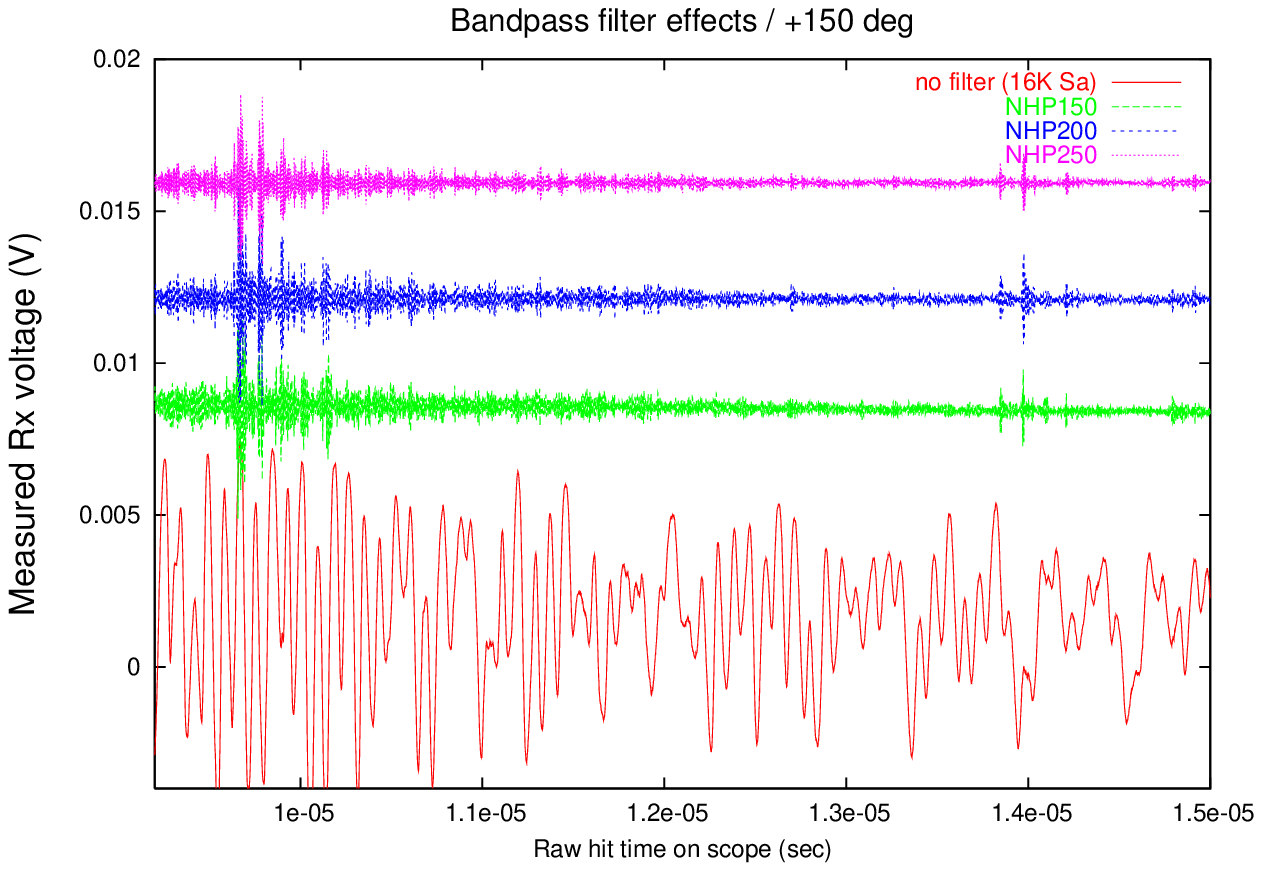}
\caption{Filter comparison, zooming in on 5 microseconds of data between 9.2 and 14.2 $\mu$s after trigger.}
\label{fig:9.2us-14.2us-Filter_comparison}
\end{minipage}
\hspace{1pc}
\begin{minipage}{18pc}
\includegraphics[width=7cm]{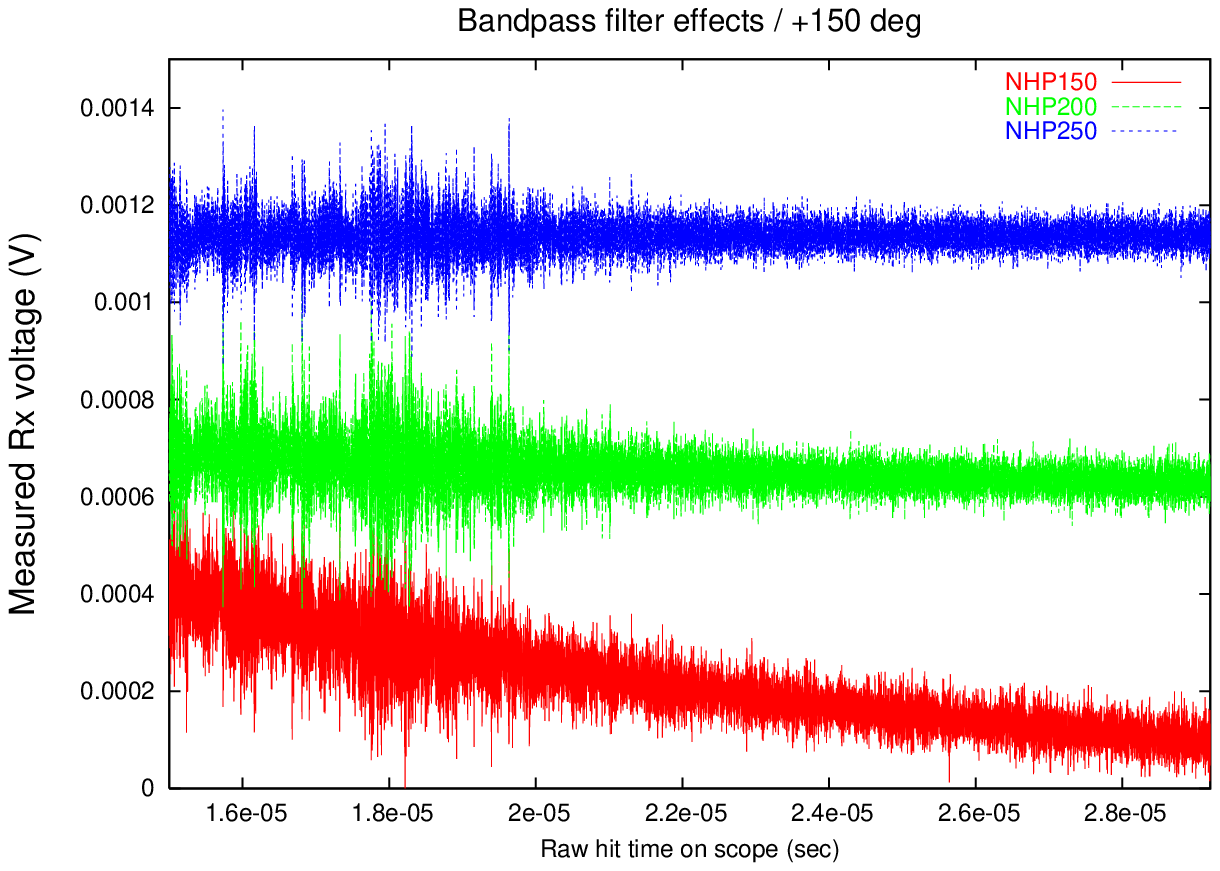}
\caption{Filter comparison, zooming in on 15 microseconds of data between 14.2 and 29.2 microseconds after trigger.}
\label{fig:14.2us-29.2us-Filter_comparison}
\end{minipage}
\end{figure}

Following the in-air measurements, the primary
measurements of reflection returns when broadcasting through 
the ice were then conducted. Despite the shielding expected from the
intervention of MAPO between transmitter and receiver, 
considerable through-air signal leakage between Tx and Rx 
is observed just after the trigger.
This large background dominates our acquired waveforms for 
approximately 5 microseconds after the initial trigger, and
prevents observation of in-ice reflections during that time
period.
In the absence of lowpass filtering, this large background will
saturate the amplifier, resulting in an amplifier settling time
which extends tens of microseconds. Most of this power, however,
is broadcast at low frequencies and can be suppressed with
appropriate filtering. Figure
\ref{fig:Filter_comparison} shows sample waveforms
acquired using different filters, beginning several microseconds
after the initial trigger. In the absence of any filtering, no signals
are clearly visible. Filtering out components below 200 MHz is sufficient to
mitigate amplifier saturation effects. Our default configuration
employs a Mini-Circuits model \#NHP250 
highpass filter, which provides $>$70 dB suppression
below 200 MHz, followed by a logarithmic turn-on through the
200-250 MHz regime.
Figures \ref{fig:9.2us-14.2us-Filter_comparison} and 
\ref{fig:14.2us-29.2us-Filter_comparison} show the waveforms obtained
for the time intervals corresponding to 9.2--14.2 and 14.2--29.2 microseconds
after the scope trigger, for filtering with NHP250 vs. NHP200 filters.
In detail, the latter still shows some indications of amplifier settling.
The gross features of the two waveforms are observed to be comparable.

\begin{figure}
\begin{minipage}{18pc}
\includegraphics[width=7cm]{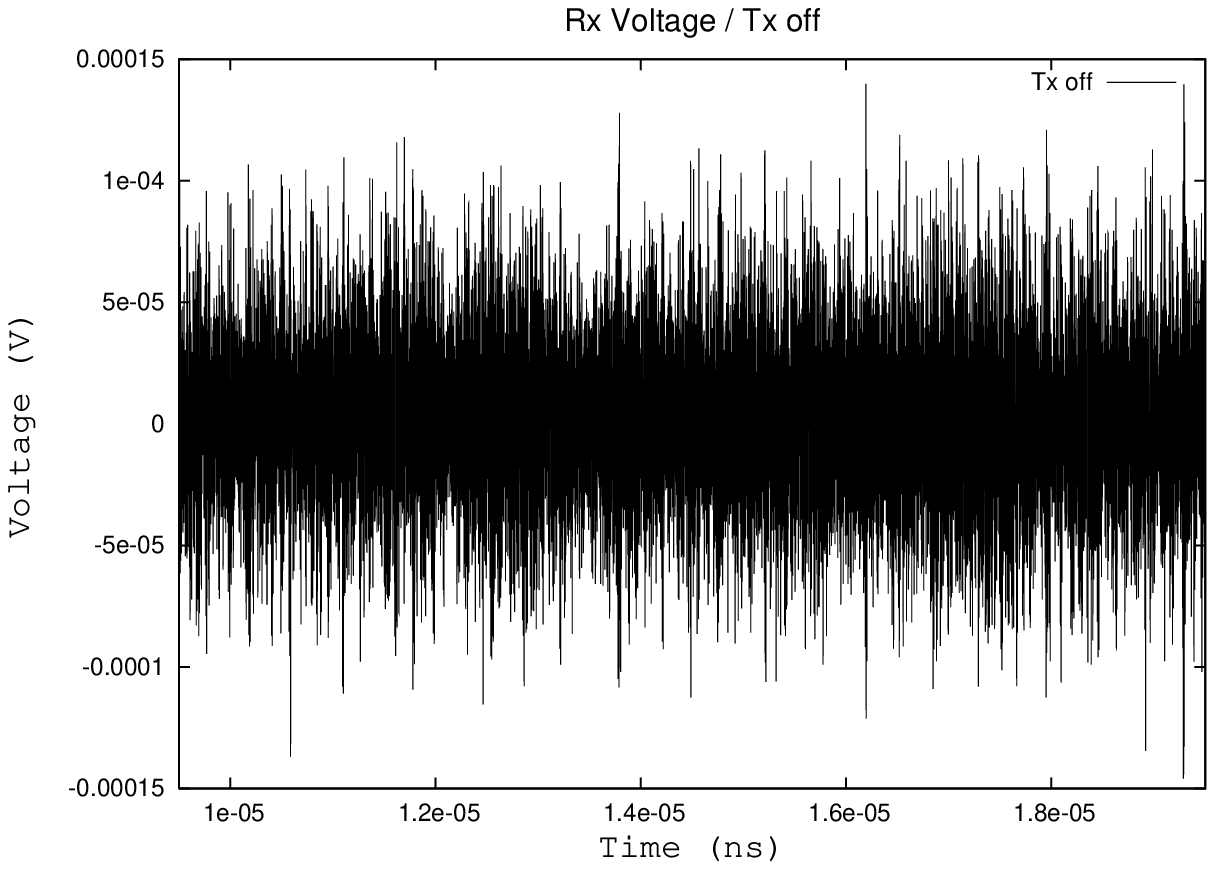}
\caption{Time-domain voltage profile, transmitter off.}
\label{fig:Txoff}
\end{minipage}
\hspace{1pc}
\begin{minipage}{18pc}
\centerline{\includegraphics[width=7cm]{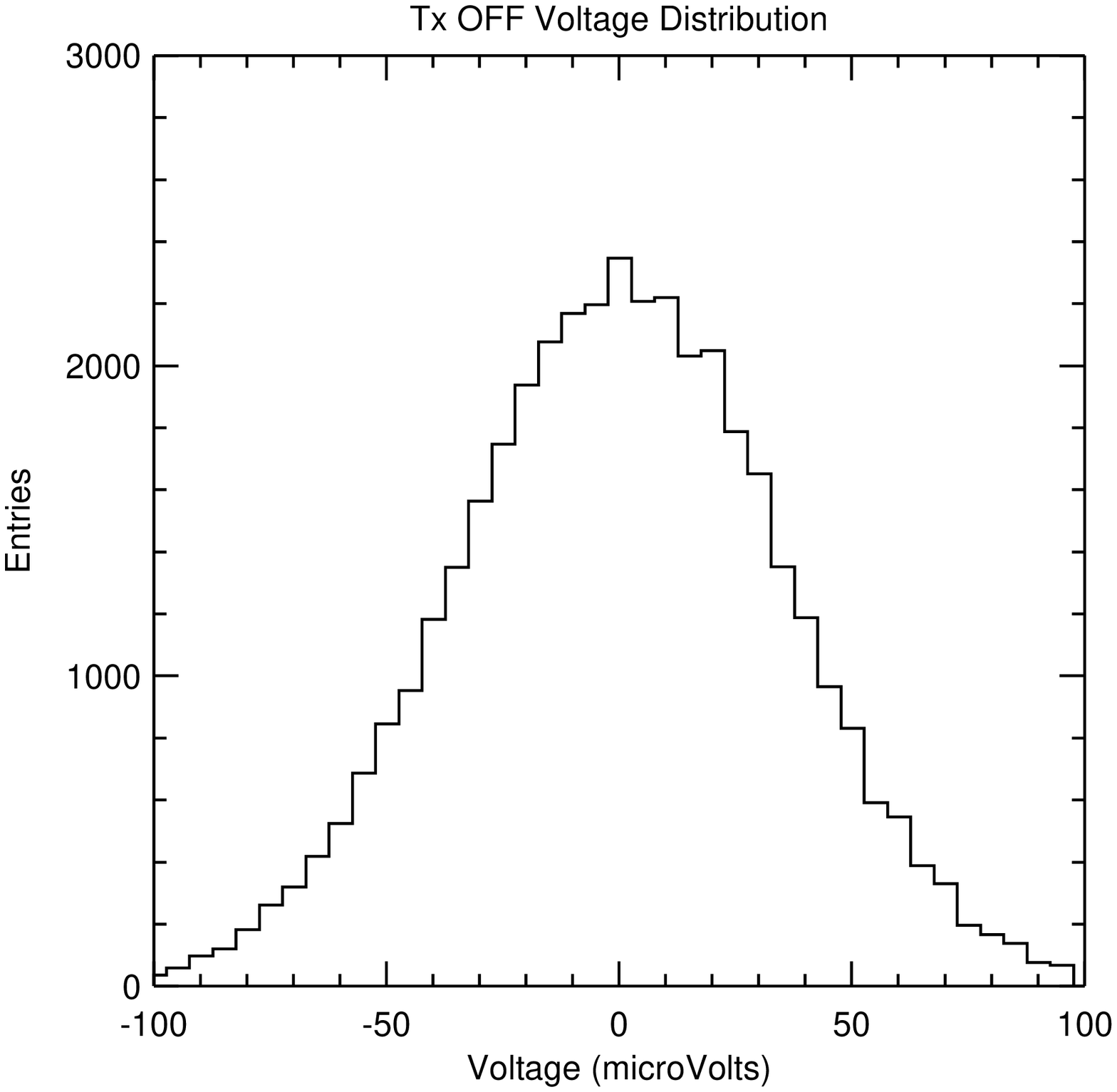}}
\caption{Projection of previous plot onto voltage axis, showing 
Gaussian-like distribution expected for thermal noise.}
\label{fig:Txoff-voltage-dist}
\end{minipage}
\end{figure}

With the NHP250 filter in place,
measurements were made to sample the noise environment,
in the absence of any transmitted signal. Figure \ref{fig:Txoff}
shows the acquired waveform. 
Figure \ref{fig:Txoff-voltage-dist}
shows the distribution of acquired voltages; we observe a roughly
Gaussian distribution, as expected by thermal noise. The rms of
this distribution is measured to be 35 microVolts. In the absence of
any noise contributed by the digital oscilloscope itself, this voltage
can be compared to the thermal noise voltage at T=223 K expected in a
$\sim$750 MHz bandpass $B$ looking into an
$R$=50 $\Omega$ load incoherently averaged over
N=40000 samples: $V\sim\sqrt{kTBR/N}$. Taking into account the cable
losses already shown, as well as the
effect of the +52 dB amplifier and the filter, we obtain an
estimate 
of approximately 30 microvolts, roughly consistent with observation.

\begin{figure}
\centerline{\includegraphics[width=7cm]{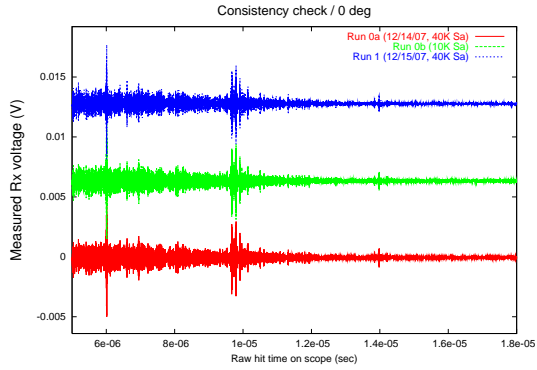}}
\caption{Check of consistency between measured waveforms, all taken in $\phi$=0 orientation.}
\label{fig:Consistency_check_parMAPOruns}
\end{figure}

In our initial configuration, both antennas were co-aligned with
their long axis parallel to the MAPO building (this simply provided
an easily reproducible
$\phi$=0 azimuthal reference and is identical to the $\phi$=0 line
shown in Figure 1.). 
Figure \ref{fig:Consistency_check_parMAPOruns} shows that: a) waveforms are reproducible from one 40K average to
the next, and b) after moving the antennas around for one day, and then
replacing them at their original location, waveforms are also reproducible.

\begin{figure}
\centerline{\includegraphics[width=7cm]{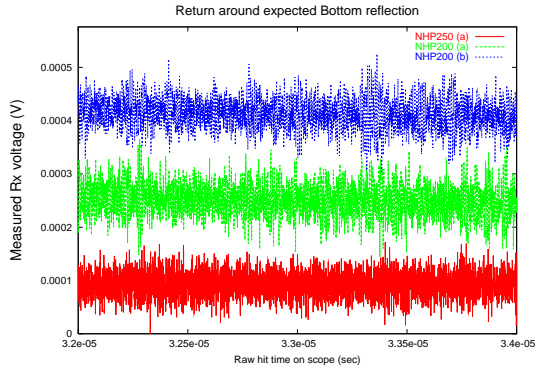}}
\caption{Return signal in time interval expected to bracket bedrock reflection signal.}
\label{fig:Bottom_bounce_comparison}
\end{figure}
\begin{figure}
\begin{minipage}{18pc}
\includegraphics[width=7cm]{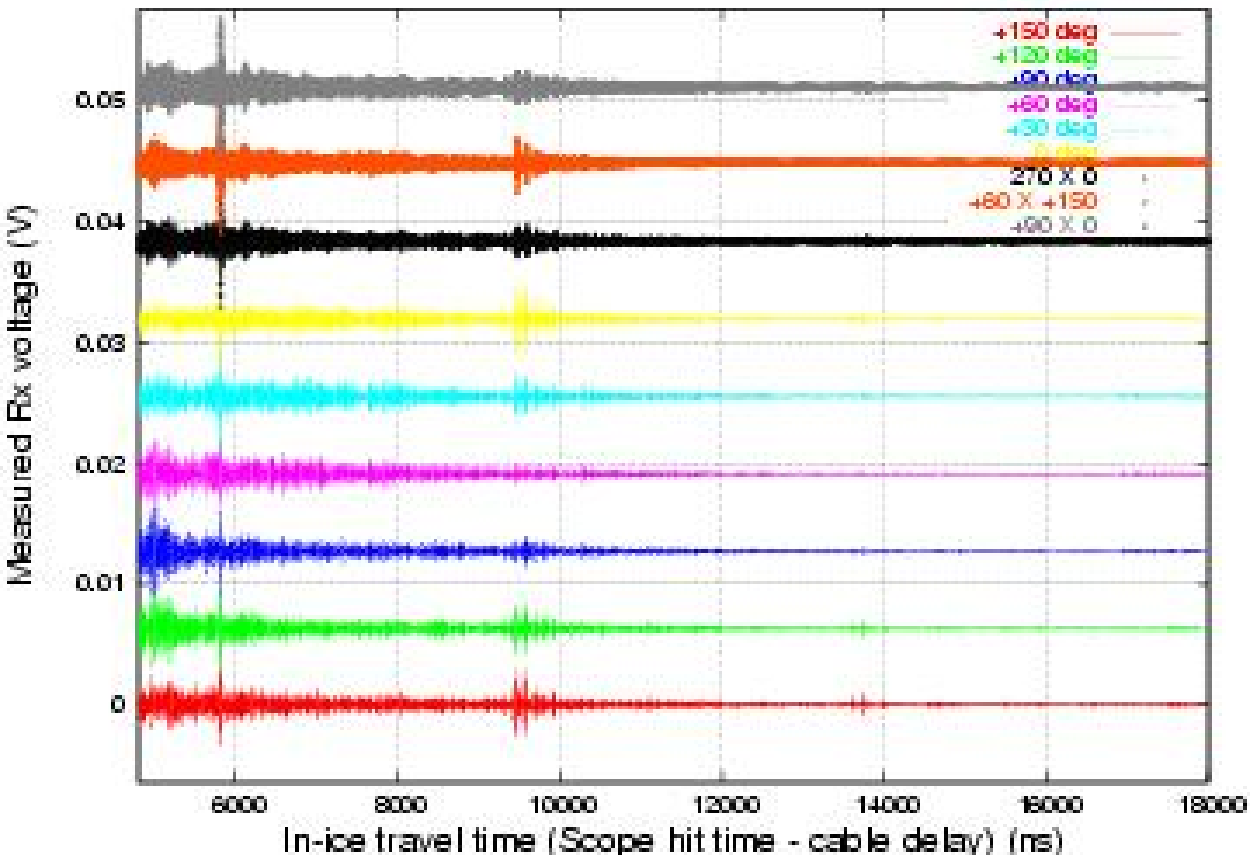}
\caption{5--25 microsecond waveform captures, after averaging, for various azimuthal angles. Data configuration corresponds to default for most of measurements cited in this paper.}
\label{fig:5us-25us-all-vertically-shifted}
\end{minipage}
\hspace{1pc}
\begin{minipage}{18pc}
\includegraphics[width=7cm]{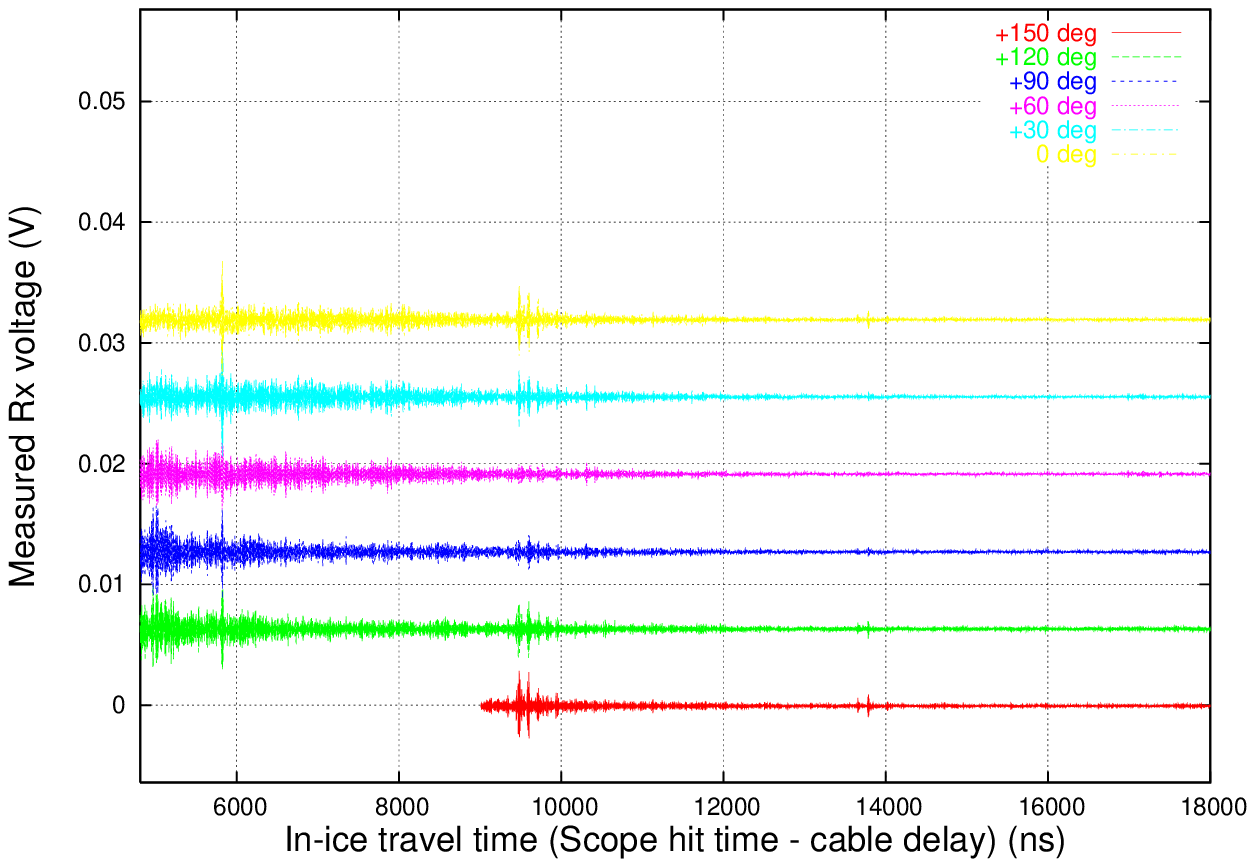}
\caption{5--25 microsecond waveform captures, after averaging, for 
various azimuthal angles. Antennas have been translated along y-axis 
(orthogonal to MAPO), but orientation angles are preserved.}
\label{fig:5us-25us-all-vertically-shifted_1}
\end{minipage}
\end{figure}

We attempted to detect birefringent effects using the same approach as
that used in our previous analysis\citep{TD-paper}, namely, we search for
a measurable time difference in received signals, as a function of
the orientation of the long axis (transmitted signal
polarization axis) of the TEM horns.
Our initial hope was to be able to detect the reflection off of the
bedrock, some 2.8 km below the surface. Due to the relative weakness
of our signal generator, however, as well as the large absorption
expected in the warmer ice near the bedrock, no clear signal
is observed in the time delay region around 33 microseconds, 
as expected for a bottom reflection (Fig. \ref{fig:Bottom_bounce_comparison}).
A follow-up measurement will employ a signal generator with a roughly
order-of-magnitude higher voltage output.
Since the bottom reflection was too weak
to be observable, internal layers were used as scattering
planes. We co-rotated the horns in the positive $\phi$
direction, and took measurements every 30 degrees. In addition, data
were taken in 3 different ``cross-polarization'' configurations, designated
by a Tx orientation and an Rx orientation.
Figure
\ref{fig:5us-25us-all-vertically-shifted} shows the signals received over
the time interval from 5--25 microseconds after the initial trigger.

As a check that the general features are relatively insensitive to the
exact placement of the antennas, data were also taken with the transmitter horn
and receiver horn both displaced approximately 10 meters in the +90 degree
direction (Figure \ref{fig:5us-25us-all-vertically-shifted_1}), and with
a slightly different trigger (and also trigger delay) configuration.
The general signal features look very similar in the displaced 
configuration as the original configuration.

\begin{figure}
\centerline{\includegraphics[width=7cm]{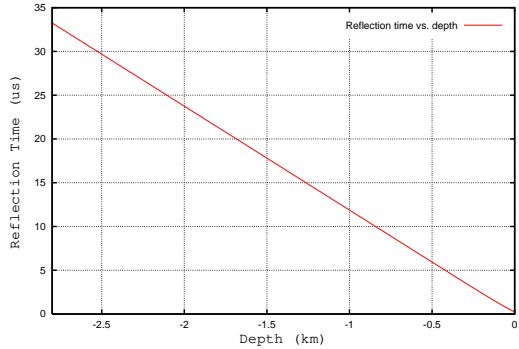}}
\caption{Expected reflection return time, as a function of depth of reflecting layer.}
\label{fig:ReflTimeVDepth}
\end{figure}

\begin{figure}
\begin{minipage}{18pc}
\includegraphics[width=7cm]{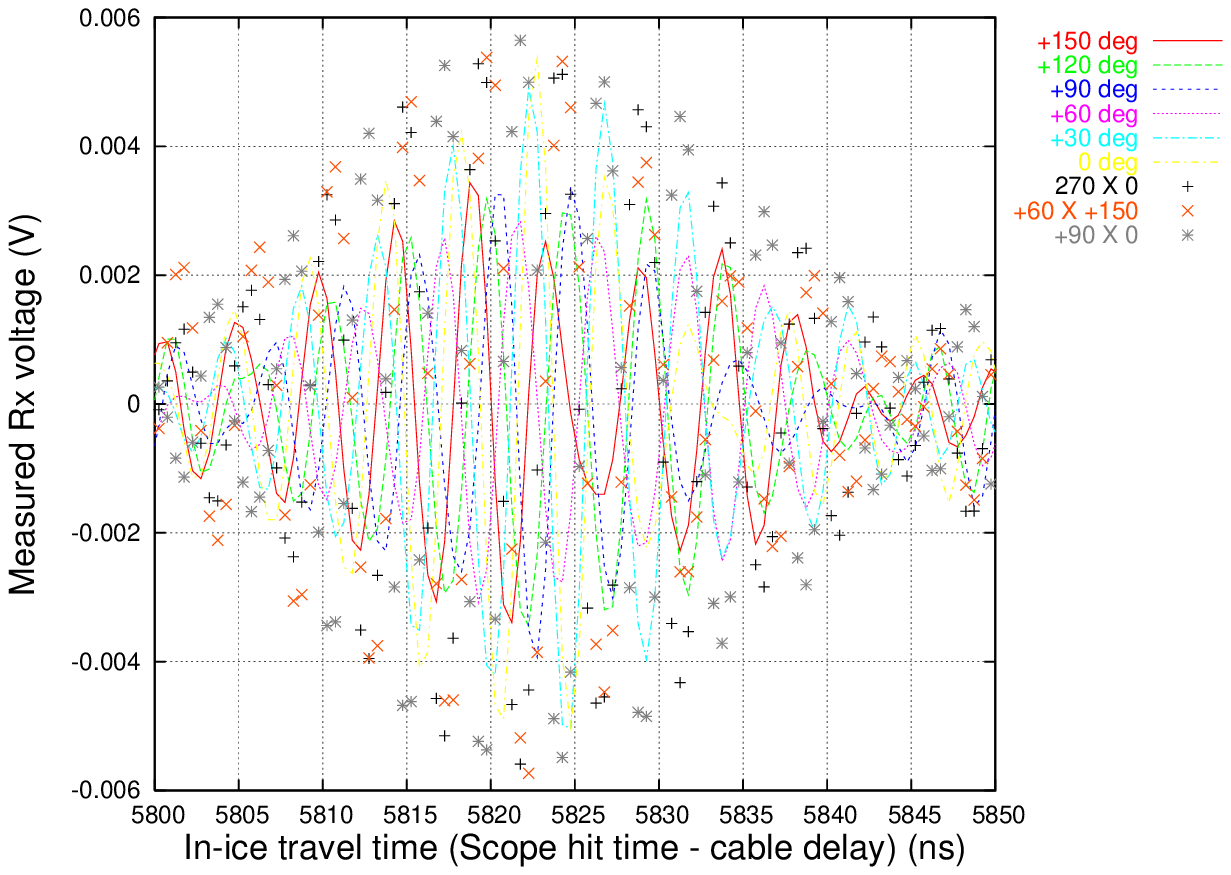}
\caption{Overlay of signal reflections around t=6 microseconds after trigger.}
\label{fig:6us-all-signal-overlay}
\end{minipage}
\hspace{1pc}
\begin{minipage}{18pc}
\includegraphics[width=7cm]{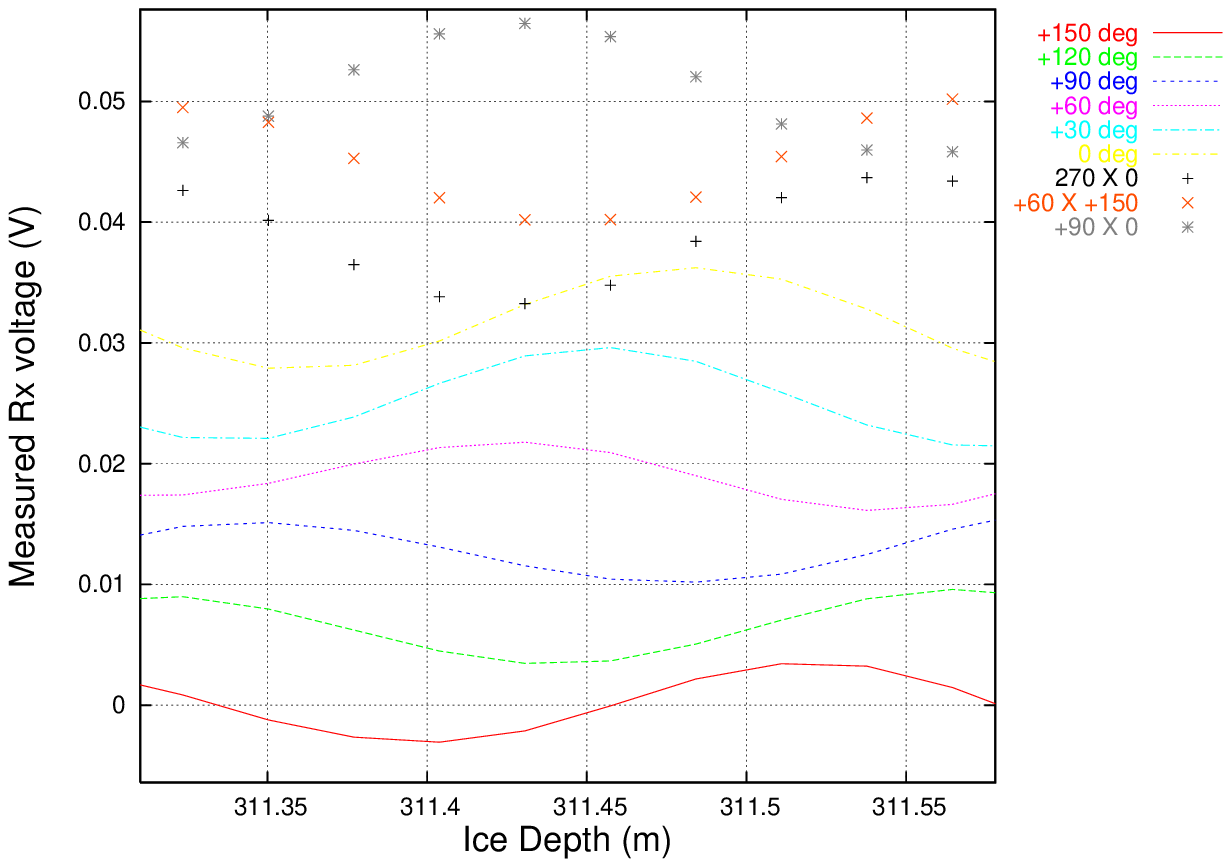}
\caption{Zoom of previous figure, showing monotonic variation in phase as a function of orientation angle.}
\label{fig:6us-zoom-zoom-vertically-shifted}
\end{minipage}
\end{figure}

The first clear return is observed at a time delay of approximately 6 microseconds
after the trigger time.
Knowing the index-of-refraction profile at South Pole,
we convert this to return signal vs. 
depth (Figure \ref{fig:ReflTimeVDepth}).
We observe a small, apparently monotonic variation in the received signal
phase with $\phi$ (Figure \ref{fig:6us-zoom-zoom-vertically-shifted}). 
Note that a value of 0.12\% birefringence, as measured in our
Taylor Dome analysis, would imply a shift of amplitude $\sim$4 ns between
Tx/Rx alignment with the ordinary vs. extraordinary optical axes. The
observed maximum time shift ($\sim$2 ns) implies a birefringent asymmetry
of $<$0.05\%. We cannot exclude the possibility that some of the observed
phase shifts are the result of interference effects between signals with
arrival times coincident within $\sim$2 ns.
We also observe a considerable anisotropy in the received signal
amplitude, which also varies sinusoidally. This could be
explained as the result of a polarization-dependent reflection coefficient.
It might also be explained as the result of polarization-dependent
absorption.

\begin{figure}
\begin{minipage}{18pc}
\includegraphics[width=7cm]{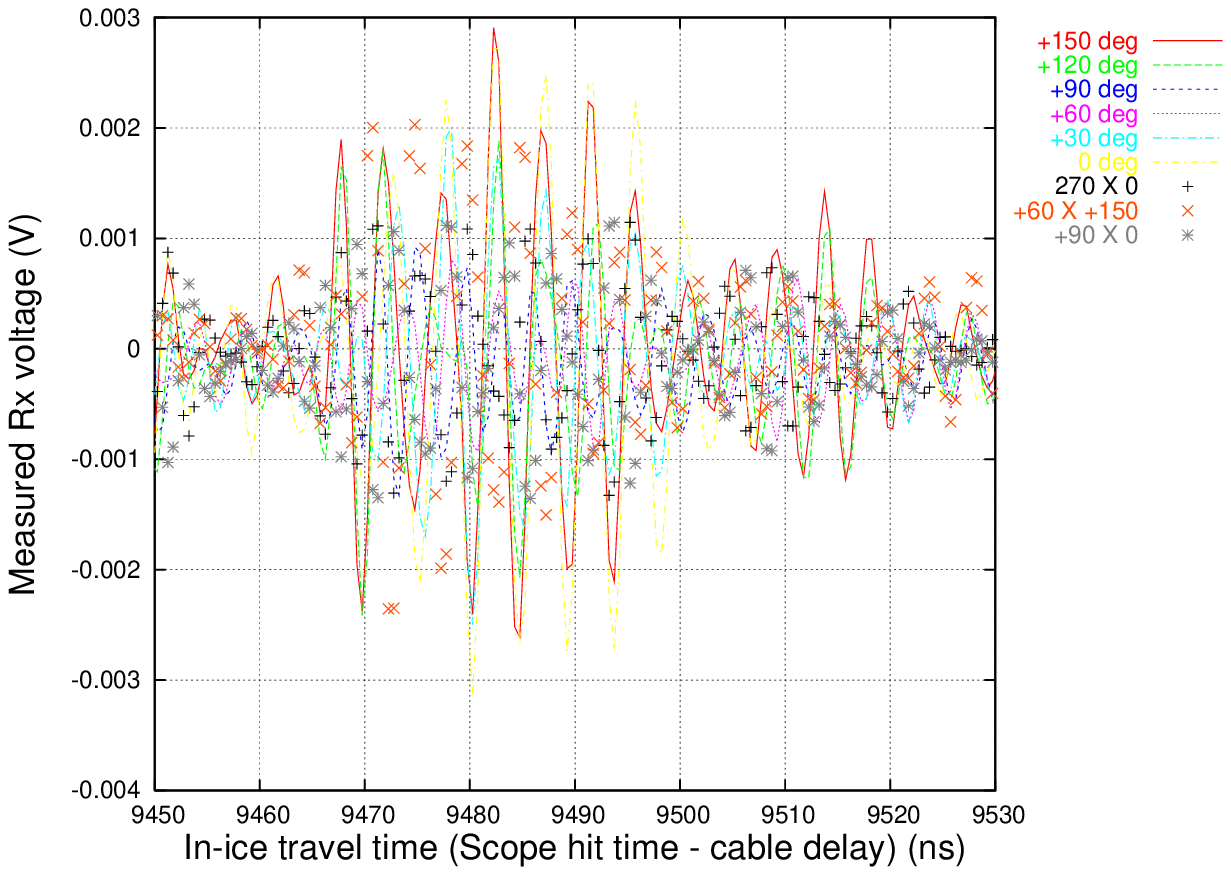}
\caption{Ensemble of observed reflections in time interval around 9.6 microseconds after trigger.}
\label{fig:9.6us-all}
\end{minipage}
\hspace{1pc}
\begin{minipage}{18pc}
\centerline{\includegraphics[width=7cm]{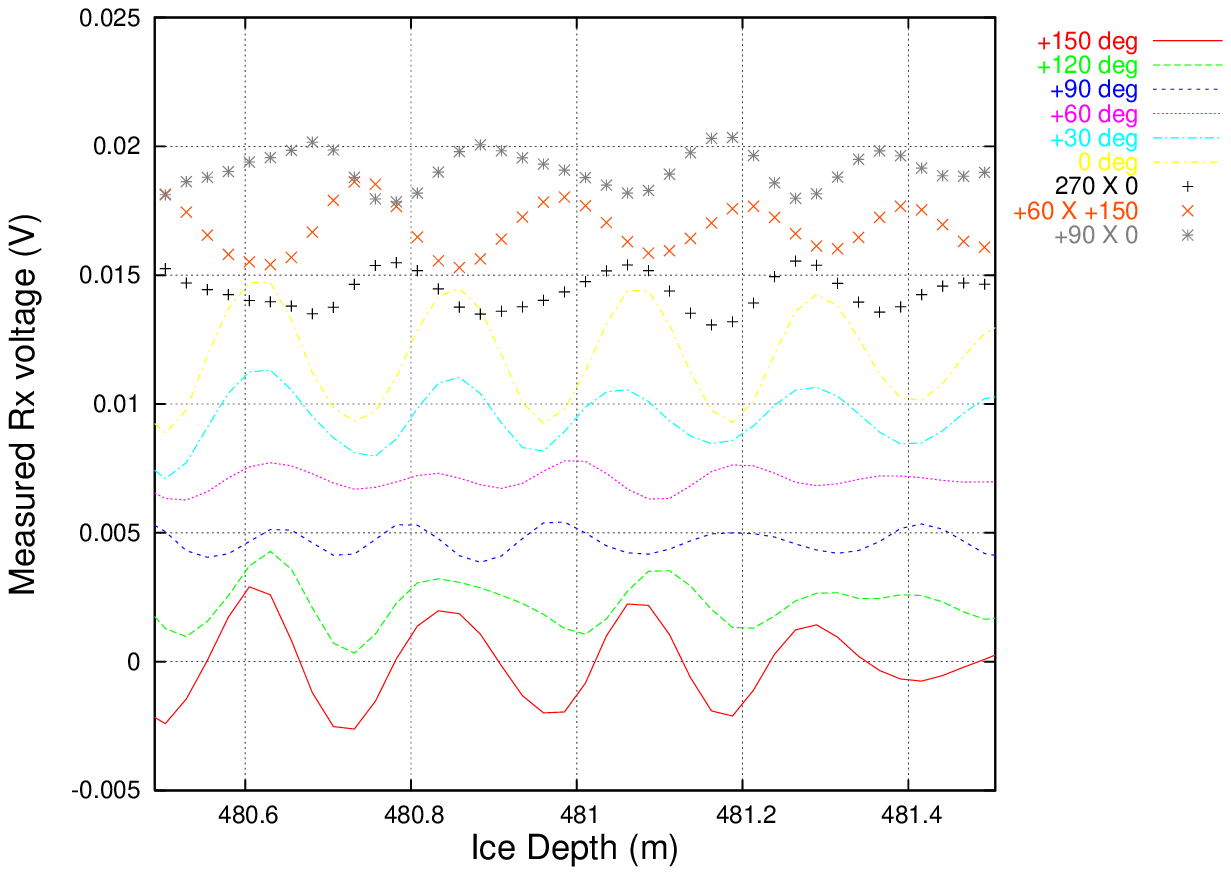}}
\caption{Zoom of observed reflections in time interval around 9.6 microseconds after trigger.}
\label{fig:9.6us-zoom}
\end{minipage}
\end{figure}

\begin{figure}
\centerline{\includegraphics[width=10cm]{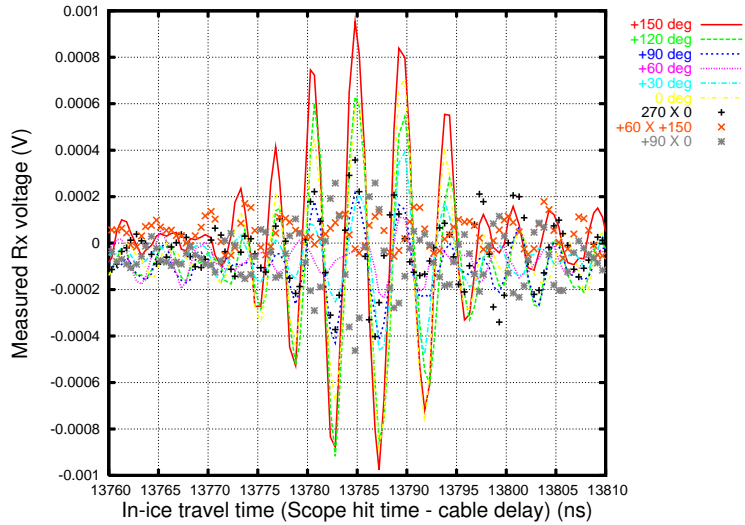}}
\caption{Overlay of 13.9 microsecond reflections, as a function of
orientation angle, default geometry and trigger configuration.}
\label{fig:13.9us-all}
\end{figure}

\begin{figure}
\centerline{\includegraphics[width=9.5cm]{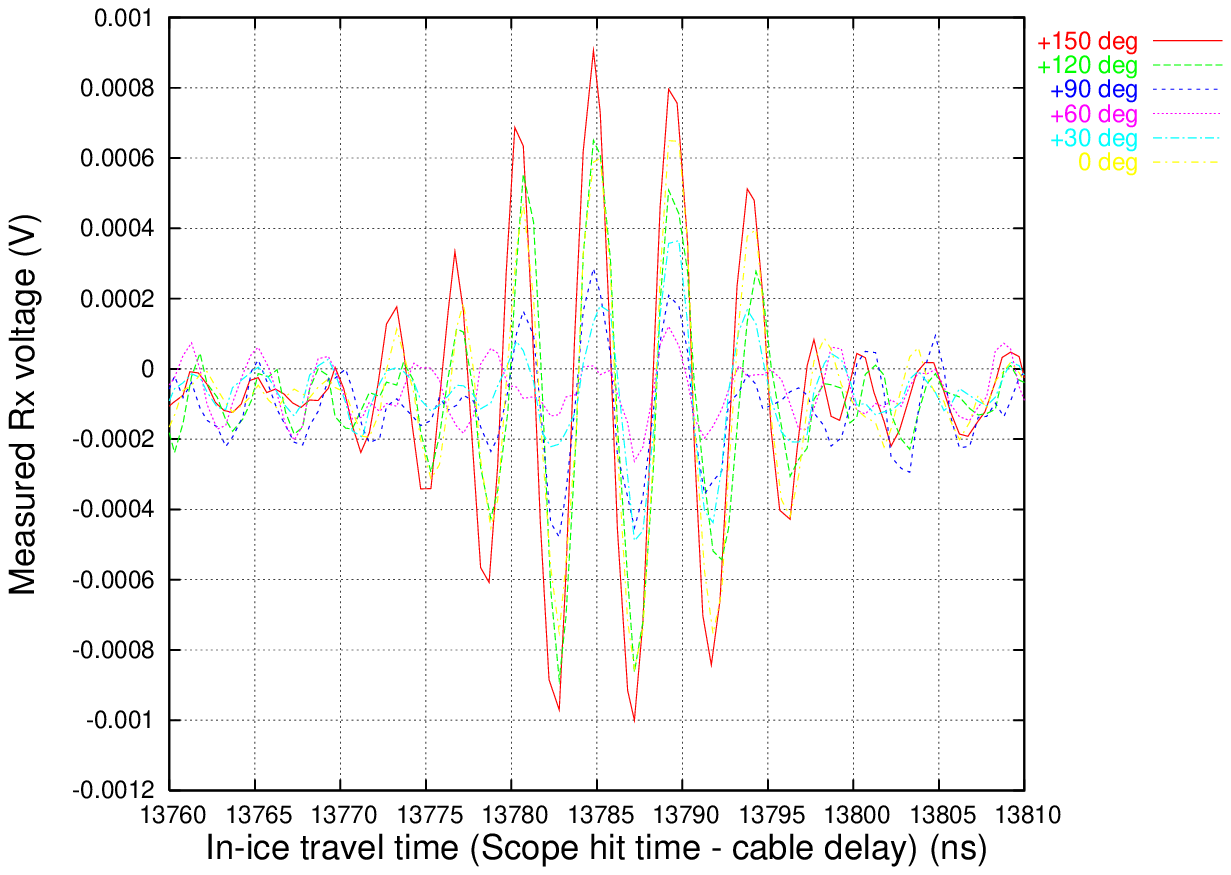}}
\caption{Similar to previous plot, but made with antennas translated on
the snow surface, and with a slightly different trigger configuration.}
\label{fig:13.9us-all_1}
\end{figure}

\begin{figure}
\centerline{\includegraphics[width=7cm]{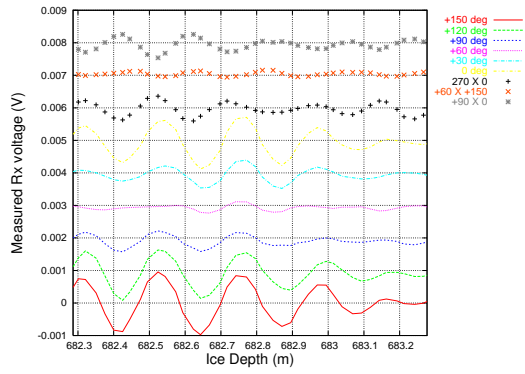}}
\caption{Zoom of 13.9 microsecond reflections, default configuration.}
\label{fig:13.9us-zoom}
\end{figure}

\begin{figure}
\centerline{\includegraphics[width=7cm]{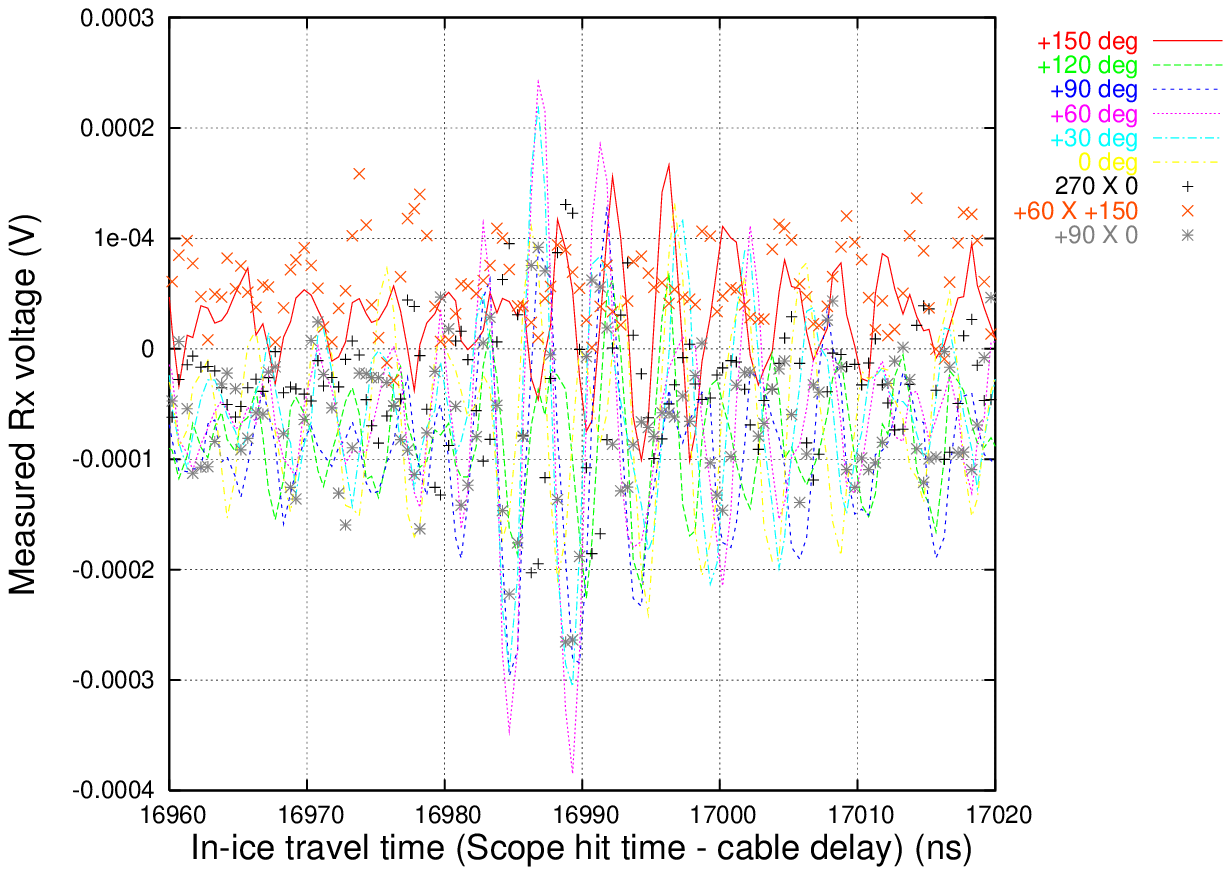}}
\caption{Ensemble of reflections observed in time interval around 17.2 microseconds after trigger.}
\label{fig:17.2us-all}
\end{figure}

\begin{figure}
\centerline{\includegraphics[width=7cm]{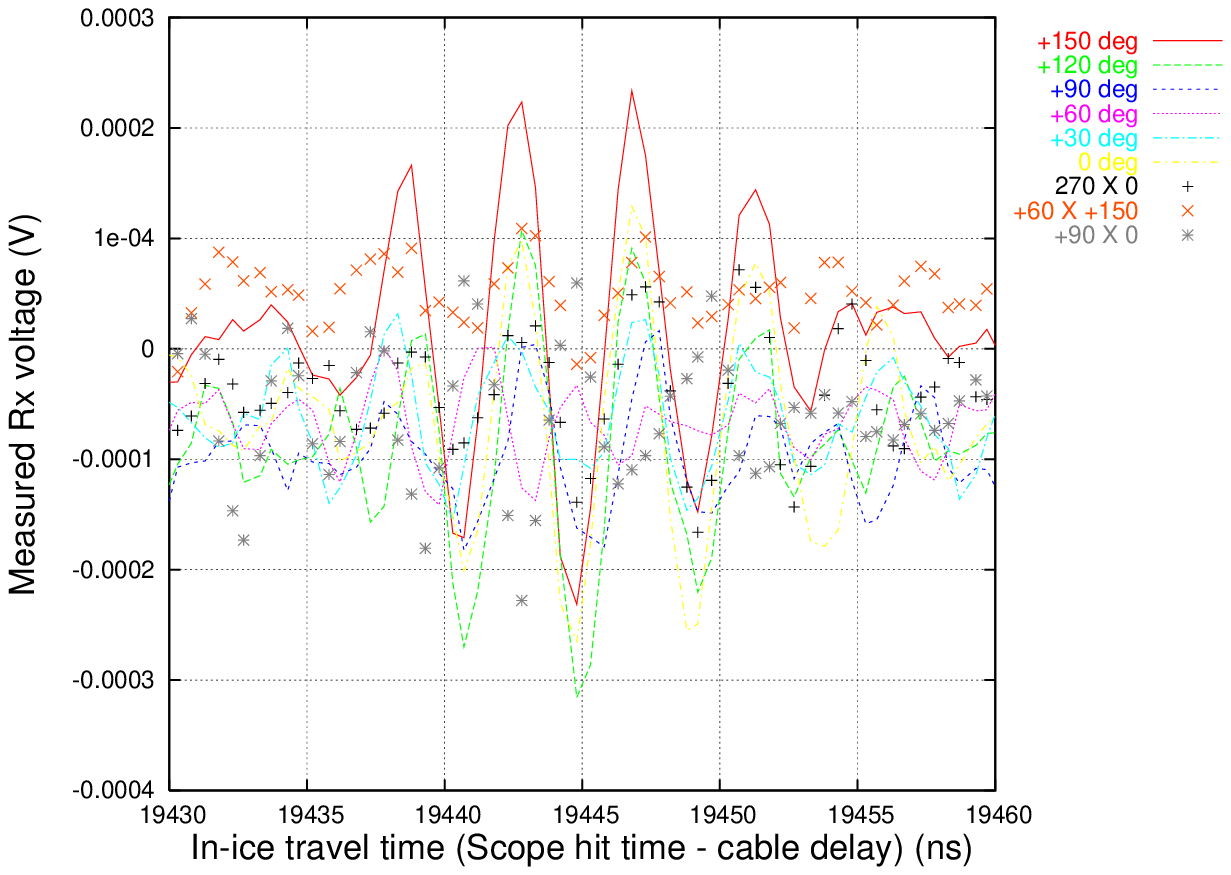}}
\caption{Ensemble of reflections observed in time interval around 19.6 microseconds after trigger.}
\label{fig:19.6us-all}
\end{figure}

Our resolution on birefringent effects obviously increases with 
the depth of the layer probed. Starting with
Figure \ref{fig:5us-25us-all-vertically-shifted}, we see clear
subsequent
structure at time delays of 9.6, 13.9, 17.2, and 19.6 microseconds, as
zoomed in Figures \ref{fig:9.6us-all}, 
\ref{fig:9.6us-zoom},
\ref{fig:13.9us-all}, 
\ref{fig:13.9us-all_1} (slightly different geometry)
\ref{fig:13.9us-zoom},
\ref{fig:17.2us-all}
and \ref{fig:19.6us-all}.

We select the 13.9 microsecond delayed signals as those with
sufficient signal-to-noise, yet with a longer delay than the 
Taylor Dome results as our primary dataset. We note in 
Figure \ref{fig:13.9us-all} that the observed waveforms 
are very consistent with each other in both shape and 
arrival time. Assuming that we
are sensitive to shifts of order $\pi/2$ (i.e., $\approx$1 ns), 
this puts a limit on the
birefringent asymmetry at South Pole to be $<$0.01\%. 
As a check of this conclusion,
we have compared the signals observed in the primary data-taking
configuration with signals observed with displaced antennas and 
slightly different trigger (Figure \ref{fig:13.9us-all_1}). The secondary
data configuration gives results consistent with the primary data
configuration.

The 19.6 microsecond data has
the largest time delay of our observed signals.
Although there is an apparent shift of a few
ns between the +150 degree and +60 degree peaks, 
the relatively larger noise contribution
to the latter waveform renders this comparison somewhat inconclusive.
All other time delays, relative to the +150 degree data, are consistent with
zero, and consistent with the interpretation of no clearly observable
birefringent effects.

\subsection{Comment on Reflected Signal Strength}
For coherent scattering,
the strength of any reflected signal is expected to follow the radar
equation for the received power 
$P_{Rx}=P_{Tx}{\tt R}G_{Tx}G_{Rx}e^{-r/L_{atten}}/(4\pi r^2)$,
with $G_{Tx}$ and $G_{Rx}$ the forward gain of transmitter and
receiver in-ice, $r$ the total transit distance (i.e., twice the depth),
$L_{atten}$ the power
attenuation length over the frequency range of interest,
and ${\tt R}$ the power reflection coefficient. 
Alternately, $G_{Rx}$ is often equated to the radar cross-section
$\sigma$
times the effective area $A_{eff}$ of the receiving antenna. We 
estimate this
product to be 
equal to 0.5 m x 0.5 m, and $\sigma\sim 4\pi~m^2$.
Given a value of $L_{atten}$,
one can therefore estimate the reflection coefficient based on the
measured power. In practice, the uncertainty on the attenuation length
is sufficiently large that this error dominates any estimate of ${\tt R}$.
As expected,
the most prominent reflections are those due to the closest layers,
corresponding to the 
most shallow returns.
As before,
we do observe a large variation in the amplitude of the return signal for
the 13.9 us reflected signal (``$A_{13.9}$),
however (Table \ref{t:amp-comparison}). 
We also observe considerable power evidently transferred from
the co-pol to the cross-pol orientation.
This will require more detailed investigation in the future
in order to assess the possible impact on neutrino detection efforts.

\begin{figure}
\centerline{\includegraphics[width=7cm]{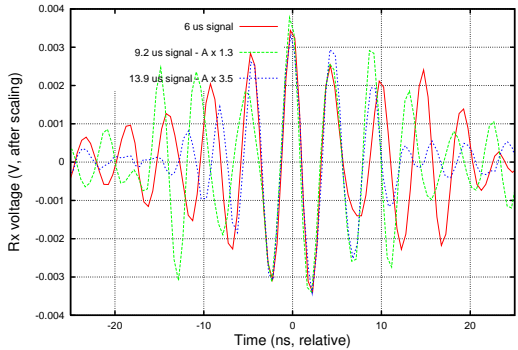}}
\caption{Comparison of shape of 6 microsecond, 9.2 microsecond, and
13.9 microsecond reflections.}
\label{fig:WF-comparison}
\end{figure}

\begin{table}
\begin{center}
\begin{tabular}{cc}
Orientation & $|V_{max}|$ (mV) \\ \hline
+180 deg & 0.897 \\
+150 deg & 0.975 \\
+120 deg & 0.917 \\
+90 deg & 0.429 \\
+60 deg & 0.235 \\
+30 deg & 0.477 \\
0 deg & 0.881 \\ \hline
\end{tabular}
\caption{Peak voltages for 13.9 us reflection signal, as a function of angle. Note the $2\pi$ periodicity in the signal amplitude. Direct application
of the radar equation, using values of attenuation length of 
order 1 km, indicates that the reflection coefficient is less than -22 dB.
We note that the maximum amplitude signal aligns with the local
global ice flow direction to within $\sim$15 degrees.}
\label{t:amp-comparison}
\end{center}
\end{table}
We have checked the uniformity of the observed reflection return waveforms,
as shown in Fig. \ref{fig:WF-comparison}. For this Figure, signals
have been scaled by the appropriate scale factor needed to match the
peak voltage to that observed at $t\sim 6\mu$s. If the reflection
coefficients were uniform, we would expect the ratios of ($A_{9.2}/A_6)/(A_{13.9}/A_{9.2}$)
to be approximately 1. Instead, we observe a ratio of ratios somewhat
less than 1, indicating non-uniformity of the reflection coefficients
at the internal scattering layers.

\begin{figure}
\centerline{\includegraphics[width=7cm]{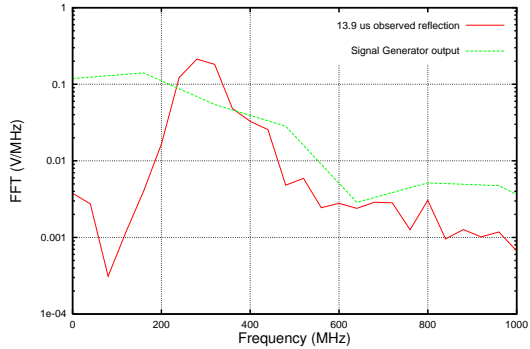}}
\caption{FFT of initially broadcast signal vs. FFT of signal measured in receiver horn.}
\label{fig:fft_hrn2hrn07.eps}
\end{figure}

In the frequency domain, the observed reflected signal retains
most of the initially broadcast signal in the frequency regime
above the highpass filter (Figure \ref{fig:fft_hrn2hrn07.eps}).
Assuming that the reflection coefficient is frequency-independent,
this is qualitatively consistent with our earlier study, which indicated
only slight dependence of attenuation length with frequency.

\section{Investigation of Faraday Rotation} 
Three obvious factors can, in principle, be responsible for
the observed cross-polarized signal strength. These
factors, which have different time dependences, are: a) cross-talk
between the H and V polarization response of the antenna (time-independent),
b) layer-specific reflection characteristics (specific at the
time of a given reflection), and c) Faraday rotation, essentially
equivalent to birefringence of circularly polarized waves. (proportional
to pathlength). Note that, since a given material experiences
a positive(/negative) phase rotation when the ${\hat k}$-vector
is parallel(/anti-parallel) to the applied magnetic field,
the phase rotation is (in general) not canceled over a
round-trip path. (This conclusion is not changed if one
assumes an
inversion of the wave when reflected off a higher index-of-refraction
medium.)
In general, although birefringent effects 
will project onto ordinary and extra-ordinary axes, it will
not result in a cross-polarized signal, although
the received signal strength will be modulated due to
interference effects, if the receiver time
resolution is sufficiently coarse.

For a given material in a magnetic field $B$,
the phase advance $\theta$ due to Faraday rotation through
a pathlength $l$ is expressed in terms of the Verdet constant
as: $\theta=VlB$. 
For a given material, the Verdet constant can be expressed
in terms of the Bohr magneton as $V=\lambda dn/d\lambda$,
with $n$ the index-of-refraction of the material. Using
$dn/d\lambda\sim 10^{-4}$ over the wavelength 
interval of interest\citep{Warren84},
we obtain (mks units), $V$=0.02 rad/(m-T)
at 300 MHz, in which case this is unlikely to be a significant effect. 

\begin{figure}
\centerline{\includegraphics[width=7cm]{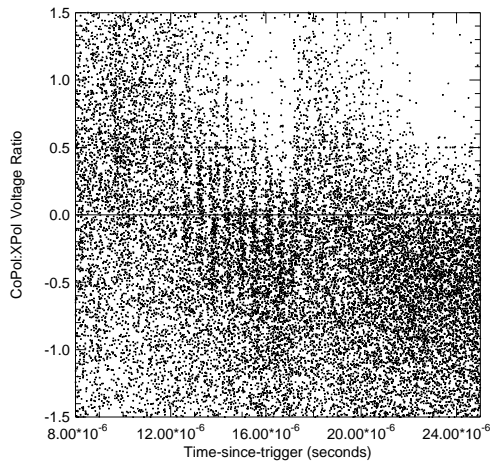}}
\caption{Ratio of co-polarized to cross-polarized signal amplitude, as a function
of time delay. Random thermal noise should average to zero on this plot; 
Faraday rotation would result in a modulated signal with time delay.}
\label{fig:Vratio}
\end{figure}

Due to its dependence on pathlength,
Faraday rotation can therefore be directly probed by plotting the
ratio of the cross-polarized amplitude to the co-polarized
amplitude as a function of time delay. Although the phase of the
cross-polarized signal observed reflecting off discrete internal layers
shows no obvious pattern (e.g., the 13.9 microsecond (/17.2 microsecond) 
+90 degree
co-polarized reflected
signal is in phase (/out of phase) with the cross-polarized signal),
the ratio of amplitudes (Figure \ref{fig:Vratio}) indicates a suggestive
modulation with period $\sim$8 microseconds. Interpreted as pure
Faraday rotation, this would indicate a $\sim\pi$ 
polarization axis rotation for each 800 m of traversed ice depth.

\begin{figure}
\centerline{\includegraphics[width=7cm]{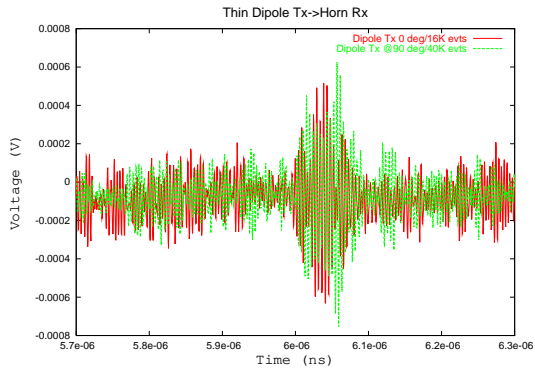}}
\caption{Signals observed using a long-dipole as transmitter antenna.}
\label{fig:greenTx2hrn}
\end{figure}

As a cross-check of the apparently large
power being fed from one polarization into the
cross-polarization, we also attempted to broadcast using 
a long dipole transmitter instead of the horn Tx, aligned
along two orthogonal axes (corresponding to zero and 90
degrees, respectively). Given the lower forward gain of the
dipoles, and the poorer frequency-response matching, 
we observe only the return at $\sim 6\mu$s, with approximately
equal amplitude (Figure \ref{fig:greenTx2hrn}) in co-pol
vs. cross-pol orientations.

\section{Comparison with previous data}
\begin{figure}
\begin{minipage}{18pc}
\includegraphics[width=7cm]{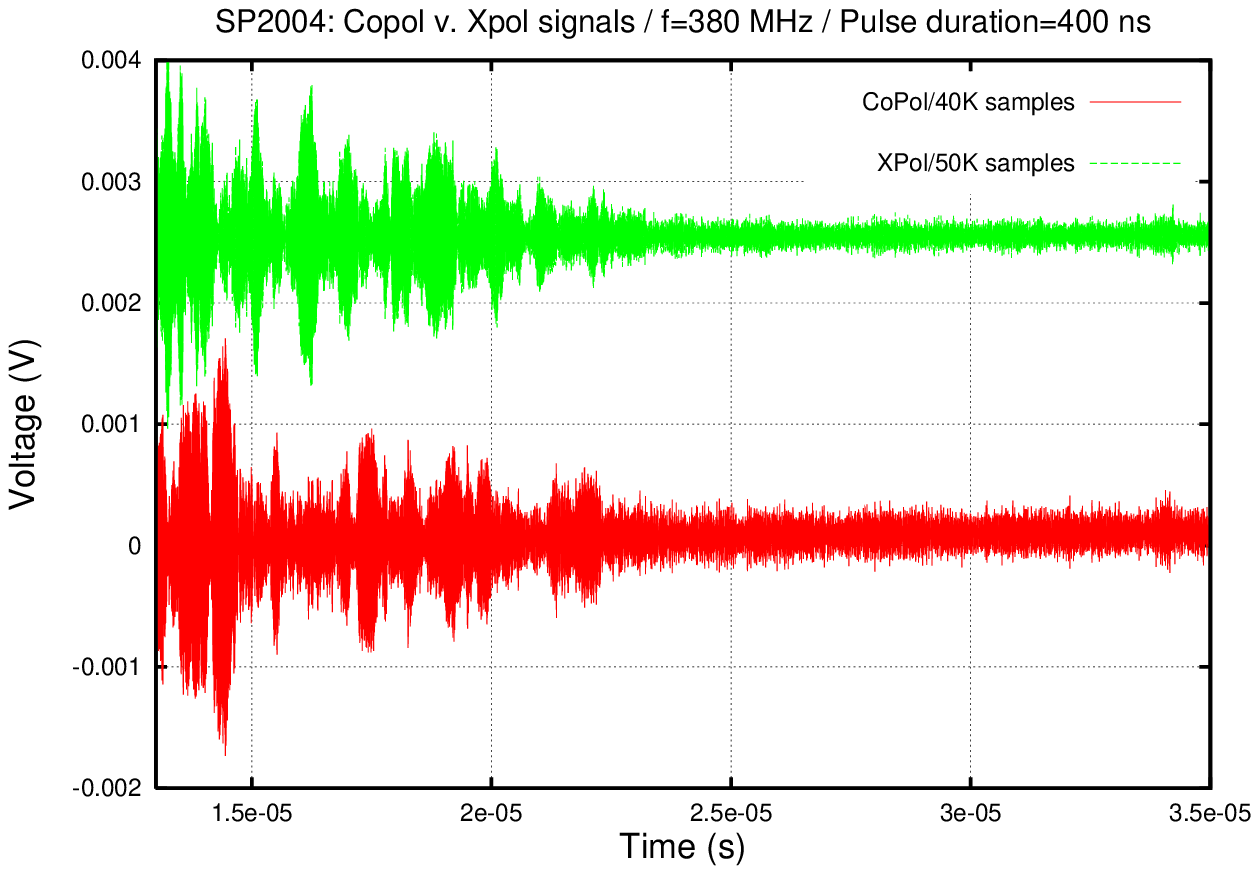}
\caption{Cross- vs. copolarized signals observed in previous study, 400 ns signal duration.}
\label{fig:2004-copol-xpol}
\end{minipage}
\hspace{1pc}
\begin{minipage}{18pc}
\includegraphics[width=7cm]{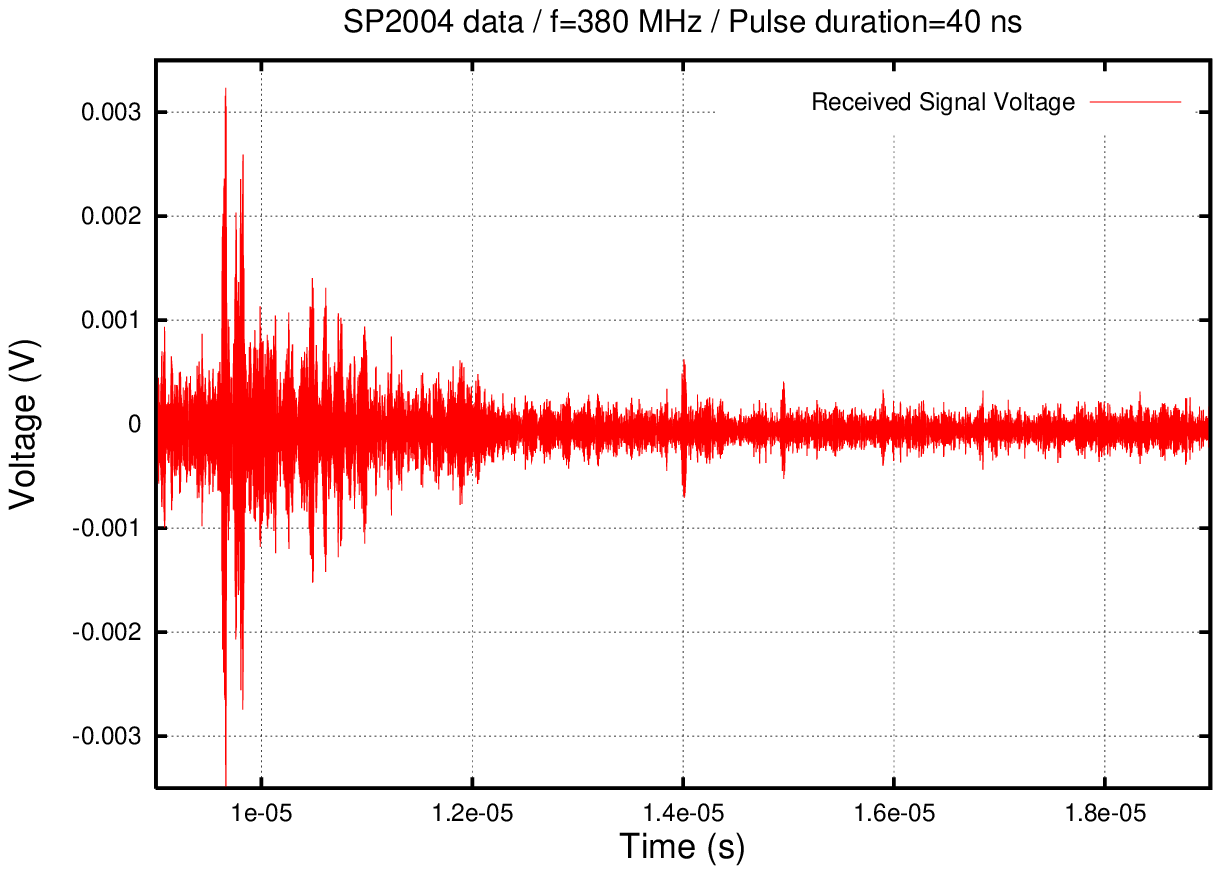}
\caption{Copol signals observed in previous study, 40 ns signal duration.}
\label{fig:2004b}
\end{minipage}
\end{figure}

Figure \ref{fig:2004-copol-xpol} shows the co-polarized vs. cross-polarized
data taken previously, in January 2004\citep{RF-eps-im}. 
Although not noted at the time,
the clear presence of the scattering layer around 14 $\mu$s is evident
from the Figures, as well as the presence of large amplitude signals in the
cross-polarization configuration, 
not necessarily present in the co-polarization configuration. The horn
orientation is approximately 160 degrees, in the coordinate system
used for our current measurement.

We note that,
in our previous study conducted
at Taylor Dome, although direct evidence was observed for birefringence in 
scattering of radio waves off the bedrock, no clear
returns from internal layers, prior to the bedrock reflection, were evident.

\begin{figure}
\centerline{\includegraphics[width=7cm]{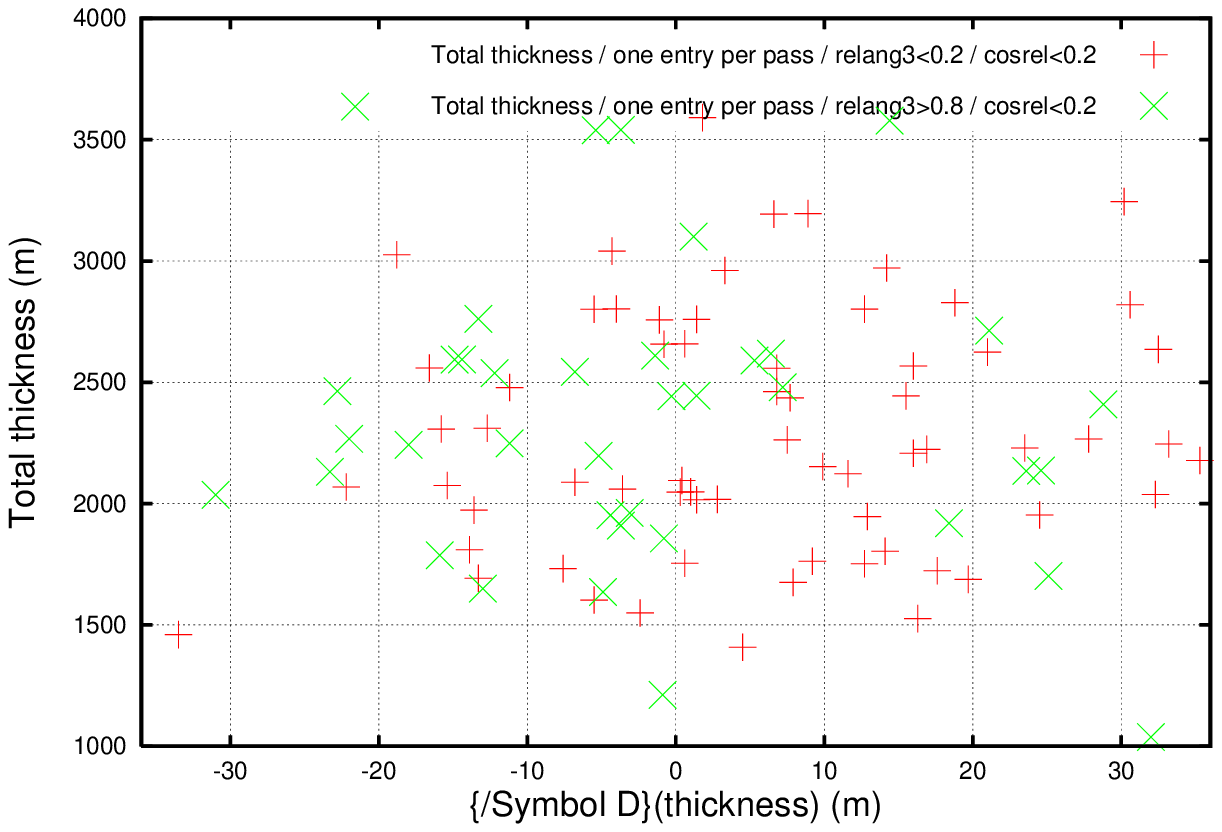}}
\caption{Depth difference recorded by CRESIS data, at a given point on the Greenland
continent. Green corresponds to difference in recorded depths for cases where plane
is flying perpendicular to known ice-flow direction vs. parallel to known ice-flow
direction; red corresponds to (parallel-perpendicular) case. Birefringence would
result in a difference in the centroid of the red vs. green points.}
\label{fig:grz-1997.eps}
\end{figure}

Using publicly available data from the CRESIS group, based at the University
of Kansas, we have also attempted to extract evidence for birefringent effects
based on their Greenland
ice thickness aerial survey data. Since the CRESIS data includes the 
aerial velocity vector as data are taken, we have searched for a statistical
separation between the depth recorded when the plane is flying parallel to the known
ice-flow direction (taken from \citep{ice-flow-direction}) vs. perpendicular to
the known ice-flow direction. Figure \ref{fig:grz-1997.eps} shows the results of
this study. The rms of these distributions ($\sim$10 m, or $\sim$100 ns in 
total in-ice travel time) is insufficient to discern clear evidence for 
the sought-after birefringent asymmetry.

\section{Summary}
There has been increasing attention given to birefringence
as responsible for 
radio-frequency interference effects observed in Antarctic ice.
Having made a similar measurement at Taylor Dome,
our goal in this analysis was to 
directly measure birefringence, as
indicated by a time-delay between
reflected signals received along two orthogonal
azimuthal polarizations. For the purposes of this study, internal
scattering
layers were used as reflecting planes, which offered the possibility
of correlating time-delays between successive reflecting planes with
the crystal orientation in those intermediate layers.
No obvious evidence for birefringence is observed. If true, this 
conclusion supports the suitability of the South Pole as a radio-based
neutrino detection site,
compared to other locales across the Antarctic
continent, where birefringent effects have been claimed. We do
observe a variation in scattered signal, as a function of azimuth,
as previously detected in a similar, albeit frequency-based, measurement.
Our results, however, disfavor that amplitude variation as a
birefringent-related effect, and may require further investigation
of the chemical and electrical properties of internal scattering layers.
There is, in fact, some data available on dust layers at the South 
Pole\citep{dust-layers}, 
however we have not made a conclusive correlation between layering
observed in this study and layers observed in previous studies.

\section{Acknowledgments}
This work is supported by the NSF through grant OPP-0338219.
The author particularly 
thanks Chris Allen (U. of Kansas), Prasad Gogineni (U. of Kansas), 
Steve Barwick (U. of California, Irvine),
Kenichi Matsuoka (U. of Washington), and
Jiwoo Nam (National Taiwan University, Taiwan)
for very helpful discussions.

\end{document}